\documentclass[default,iicol]{sn-jnl}

\usepackage[utf8]{inputenc}
\usepackage[english]{babel}
\usepackage{amsmath}
\usepackage{amssymb}
\usepackage{physics}
\usepackage [autostyle, english = american]{csquotes}
\MakeOuterQuote{"}
\usepackage{bm}
\usepackage{caption}
\usepackage{subcaption}
\usepackage{lineno}
\usepackage{academicons}

\usepackage{hyperref}
\usepackage{graphicx}%
\usepackage{multirow}%
\usepackage{amsmath,amssymb,amsfonts}%
\usepackage{amsthm}%
\usepackage{mathrsfs}%
\usepackage{appendix}%
\usepackage{xcolor}%
\usepackage{textcomp}%
\usepackage{manyfoot}%
\usepackage{booktabs}%
\usepackage{algorithm}%
\usepackage{algorithmicx}%
\usepackage{algpseudocode}%
\usepackage{listings}%

\begin{document}


\title{Improving the Performance of Cryogenic Calorimeters with Nonlinear Multivariate Noise Cancellation Algorithms}
\author*[1,2]{K. J.~Vetter}\email{kenneth\_vetter@berkeley.edu}
\author[1,2]{M.~Beretta}
\author[2]{C.~Capelli}
\author[3,4]{F.~Del~Corso}
\author[1,2]{E. V.~Hansen}
\author[1,2]{R. G.~Huang}
\author[1,2]{Yu. G.~Kolomensky}
\author[5,6]{L.~Marini}
\author[7,8]{I.~Nutini}
\author[1,2]{V.~Singh}
\author[1,9]{A.~Torres}
\author[1,2]{B.~Welliver}
\author[9]{S.~Zimmermann}
\author[3,4]{and S.~Zucchelli}

\affil[1]{Department of Physics, University of California Berkeley, Berkeley, CA 94720, USA}
\affil[2]{ Nuclear Science Division, Lawrence Berkeley National Lab, Berkeley, CA 94720, USA}
\affil[3]{Dipartimento di Fisica e Astronomia, Alma Mater Studiorum -- Universit\`{a} di Bologna, Bologna I-40127, Italy}
\affil[4]{INFN -- Sezione di Bologna, Bologna I-40127, Italy}
\affil[5]{Gran Sasso Science Institute, L'Aquila I-67100, Italy}
\affil[6]{INFN -- Laboratori Nazionali del Gran Sasso, Assergi (L'Aquila) I-67100, Italy}
\affil[7]{INFN -- Sezione di Milano Bicocca, Milano I-20126, Italy}
\affil[8]{Dipartimento di Fisica, Universit\`{a} di Milano-Bicocca, Milano I-20126, Italy}
\affil[9]{Engineering Division, Lawrence Berkeley National Lab, Berkeley, CA 94720, USA}



\abstract{State-of-the-art physics experiments require high-resolution, low-noise, and low-threshold detectors to achieve competitive scientific results. However, experimental environments invariably introduce sources of noise, such as electrical interference or microphonics. The sources of this environmental noise can often be monitored by adding specially designed "auxiliary devices" (e.g. microphones, accelerometers, seismometers, magnetometers, and antennae). A model can then be constructed to predict the detector noise based on the auxiliary device information, which can then be subtracted from the true detector signal. Here, we present a multivariate noise cancellation algorithm which can be used in a variety of settings to improve the performance of detectors using multiple auxiliary devices. To validate this approach, we apply it to simulated data to remove noise due to electromagnetic interference and microphonic vibrations. We then employ the algorithm to a cryogenic light detector in the laboratory and show an improvement in the detector performance. Finally, we motivate the use of nonlinear terms to better model vibrational contributions to the noise in thermal detectors. We show a further improvement in the performance of a particular channel of the CUORE detector when using the nonlinear algorithm in combination with optimal filtering techniques.}


\maketitle

\clearpage

\section{Introduction}\label{sec:intro}

To meet increasingly demanding physical sensitivities, experiments require low threshold detectors with low noise. Despite the excellent energy resolutions and low thresholds obtained for modern detectors, the intrinsic noise limit is often far out of reach for many of these experiments. Regardless of the information carrier used (e.g. heat, light, charge), detectors are often subject to unavoidable sources of noise when deployed in the field. These may include vibrations from nearby equipment, seismic activity, and/or electronic interference from ground loops or cross-talk from other electronics used in the experiment. For example, gravitational-wave data from the Laser Interferometer Gravitational-Wave Observatory (LIGO) \cite{aasi2015advanced} and Virgo \cite{Virgo} experiments are subject to anthropogenic microphonic noise and are sensitive to transient noise during earthquakes \cite{virgoO3}. Analyses of these data are typically conducted over a range of frequencies above $\sim$20 Hz \cite{abbott2020guide}, though this cutoff could be reduced with the reduction of anthropogenic noise below 20 Hz \cite{berger}. In the case of charge-collecting detectors used in modern gamma ray spectroscopy experiments, vibrational disturbances can change the capacitance of the system, leading to microphonic noise which degrades the detector performance, including its energy resolution \cite{zimmermann2013active}. In these cases, the signals of interest can have rise times on the order of hundreds of nanoseconds \cite{GretaDAQ}, resulting in signal bandwidths that are tens of MHz wide \cite{zimmermann2013development}. In the case of low-temperature calorimetric detectors, non-negligible amounts of noise can arise from the vibrations induced by the cryogenic facility which dissipate power, thereby adding microphonic and thermal noise to the system \cite{vibnoiseCryo}. In experiments employing such detectors (e.g. \cite{cuore2022search, cuore0, cupidMo, cupid0, amore}), the relevant signal bandwidths range from tens to hundreds of Hz \cite{Cuore_Noise}. Analyses of the CUORE data have already suggested that a large component of the noise in several channels is due to vibrations \cite{thermalModel}, warranting further study of microphonic noise.

The aforementioned sources of noise can be monitored with auxiliary devices which are specially designed to measure particular sources of interest. Deploying such devices to monitor transient noise in an experiment is a common practice. For example, the LIGO and Virgo collaborations use seismometers, accelerometers, magnetometers, and other devices to monitor non-Gaussian transient noise near their detectors \cite{virgoO3, abbott2020guide}. Compared to the cost of typical detectors used in experiments today, these devices are typically affordable while providing sufficiently high fidelities to monitor the relevant noise sources. When the auxiliary data is correlated with the noise in the detectors of interest, denoising algorithms can be applied to the data. The result is a relatively low-cost way to reduce noise without major hardware interventions that can seriously impact an ongoing experiment. Noise cancellation algorithms using data from accelerometers have been demonstrated \cite{zimmermann2013development} for possible use in the GRETINA \cite{gretina} and GRETA \cite{Greta} experiments. Noise cancellation techniques were also developed and applied to data from the CUORICINO \cite{cuoricino} and CUORE-0 \cite{cuore0_results} experiments by decorrelating the noise from different detector channels \cite{cuoricino_noise, ouellet}.

Here, we implement a denoising algorithm based on a multi-input, single-output model. In Section \ref{sec:sim}, we demonstrate the algorithm on simulated data. In Section \ref{sec:berkeley_denoising}, we apply the algorithm to a cryogenic calorimeter operated as a light detector instrumented with neutron transmutation doped (NTD) germanium \cite{haller1984ntd, wang1990electrical} as our detector of interest. We use an antenna, three accelerometers, and a separate NTD vibration sensor as auxiliary devices and demonstrate that our algorithm improves both the signal-to-noise ratio (SNR) of the detector and its energy resolution. Finally, in Section \ref{sec:cuore_denoising} we expand this algorithm to include nonlinear terms and apply this nonlinear version of the algorithm to a channel of the CUORE experiment. Using accelerometers and seismometers as auxiliary devices, we again show an improvement in SNR and energy resolution of the detector.

\section{Denoising Algorithm}\label{sec:math}

The starting point of this algorithm draws heavily from the multiple-input, single-output model outlined in \cite{bendat2011random}. We begin with $n$ timestreams $x_i[t]$ digitized at a sampling frequency of $f_s$. Here $i = 0,.., n-1$ indexes the different inputs, i.e. the auxiliary devices described in Section \ref{sec:intro}. We also require that the timestream of the detector we seek to denoise, $y[t]$, be sampled at $f_s$\footnote{Technically, the common sampling frequency requirement is a simplification. Because the transfer functions are calculated in the frequency domain, if the sampling rates differ, the transfer functions can still be calculated by truncating the signal with higher the sampling rate in the frequency domain.}. The goal is to find the array of transfer functions $H_{x_iy}[t]$ which describes the multiple-input single-output system. In practice, $y[t]$ will contain noise and signals of interest for the experiment. (Henceforth, we will refer to this signal of interest as a \textit{pulse}, while ``signal'' will be reserved to refer to digital signals such as $x_i$ and $y$.) One must therefore subdivide the timestreams $x_i[t]$ and $y[t]$ into events of length $N = T f_s$, where $T$, the length of the filter, is a tunable parameter of the algorithm. The subdivision is done by triggering the raw data stream, after which we can divide the data into two subsets. In the ideal case, one subset contains only events which are free of pulses, $\{y^{(n)}[t]\}$, which we will call noise events. The other subset, $\{y^{(p)}[t]\}$, contains all of the pulses present in the data. We take the set of noise events in the detector of interest as well as the corresponding set of events $\{x^{(n)}_i[t]\}$, which occur at the same times as the elements of $\{y^{(n)}[t]\}$. We then take the discrete Fourier transform of all noise events giving us the frequency domain sets $\{Y^{(n)}[f]\}$ and $\{X^{(n)}_i[f]\}$. With these two sets, we construct the following objects:
\begin{align}
G_{yy}[f] &= \frac{2}{T}\expval{Y^{(n)*}[f] Y^{(n)}[f]} \label{eq:gyy}\\
G_{x_iy}[f] &= \frac{2}{T}\expval{X^{(n)*}[f] Y^{(n)}[f]} \label{eq:giy}\\
G_{x_ix_j}[f] &= \frac{2}{T}\expval{X^{(n)*}[f] X^{(n)}_j[f]} \label{eq:gij} 
\end{align}

The expectation values are taken over the sets of noise events from each device that occur at the same time. For all these objects, $f$ is defined within the range $[0, f_{s})$ and discretized into $N$ bins of width $\Delta f = \frac{1}{T}$. For convenience, we will drop $[f]$ hereafter and it will be implied that we are operating in the frequency domain unless otherwise noted. It is worthwhile to describe these objects in terms of familiar quantities. $G_{yy}$ is the average noise power spectrum (ANPS) of the output signal, which is a real non-negative $1 \times N$ vector. $G_{x_iy}$ is a complex $n\times N$ matrix containing the cross-spectral densities of each input with the output. $G_{x_ix_j}$ is a complex  $n \times n \times N$ 3-dimensional array. For each frequency it contains a Hermitian matrix of the cross-spectral densities between the input signals. The on-diagonal terms $G_{x_ix_i}$ comprise the ANPS of the input signals. After averaging over all noise events, the transfer functions from the inputs to the detector $H_{x_iy}$ can be calculated:
\begin{equation}\label{eq:hxy} 
 H_{x_iy} = G_{x_ix_j}^{-1} G_{x_jy}.
\end{equation}

$H_{x_iy}$ is the same size as $G_{x_iy}$ and also complex. The denoised signal, $Y'$, is the difference between the original signal and the sum of the auxiliary signals filtered with the transfer function:
\begin{equation}
 Y' = Y - H_{x_iy}^{T} X_i.
\end{equation}
This process can then be applied to pulse events to predict the noise which is correlated with the input devices during the event. Assuming that the noise in the detector of interest has components which are correlated with the auxiliaries, the result will be a denoised version of the original set of pulse events. This assumption is usually true given a careful choice of auxiliary devices used in the experiment, however in cases where the matrix $G_{x_ix_j}$ is dominated by uncorrelated noise in the auxiliary devices, then this denoising technique will not work.

One can verify the success of the algorithm by comparing the noise power spectra of the output signal and the denoised signal, $G_{yy}$ and $G_{y'y'}$. One can then proceed to analyze the denoised data, leading to more precise measurements. It is important to note that this algorithm is not a replacement for other filtering techniques such as matched filtering. Rather, this algorithm can be used first to improve the SNR, resulting in new coefficients for the matched filter which can then be applied to the data. This procedure and the implementation of many common filters including the matched filter are linear operations so the two can be combined into a single step \cite{ouellet}.

\section{Simulation of Calorimeter Data}\label{sec:sim}

To measure the efficacy of our multivariate algorithm under known conditions, we simulate a multi-input, single output system to generate mock data of a cryogenic calorimeter. This mock signal of interest contains both noise and pulses. The simulated auxiliary timestreams represent three accelerometers measuring vibrations and one antenna measuring electromagnetic interference. We first create two signals which represent the original sources of electrical and vibrational noise. The electrical source signal is a sum of many sinusoids at harmonics of 60 Hz, each with a randomized phase.\footnote{Here we are simulating electrical noise in data taken in North America, where the standard AC frequency is 60 Hz. Later, in Section \ref{sec:cuore}, we will use data taken in Europe, where the standard AC frequency is 50 Hz.} The vibrational noise is modeled as the sum of three independent sources. The first source signal is a series of sinusoids at harmonics of 1.4 Hz representing the noise from pulse tube (PT) coolers such as the Cryomech PT 415 cryocooler \cite{Cryomech}, four of which are used in the CUORE experiment. The second source signal is another sinusoid at 100 Hz, simulating a pump or fan with a particular operating frequency. The third source signal is a randomly distributed set of infrequent impulses passed through band-pass filters at three low-frequency values. The impulses represent sudden disturbances in the system, e.g. earthquakes, and the narrow-band responses represent mechanical resonances in the system that are excited by these disturbances.

\begin{figure}[!ht]
\centering
\begin{subfigure}{\linewidth}
 \includegraphics[width=\linewidth]{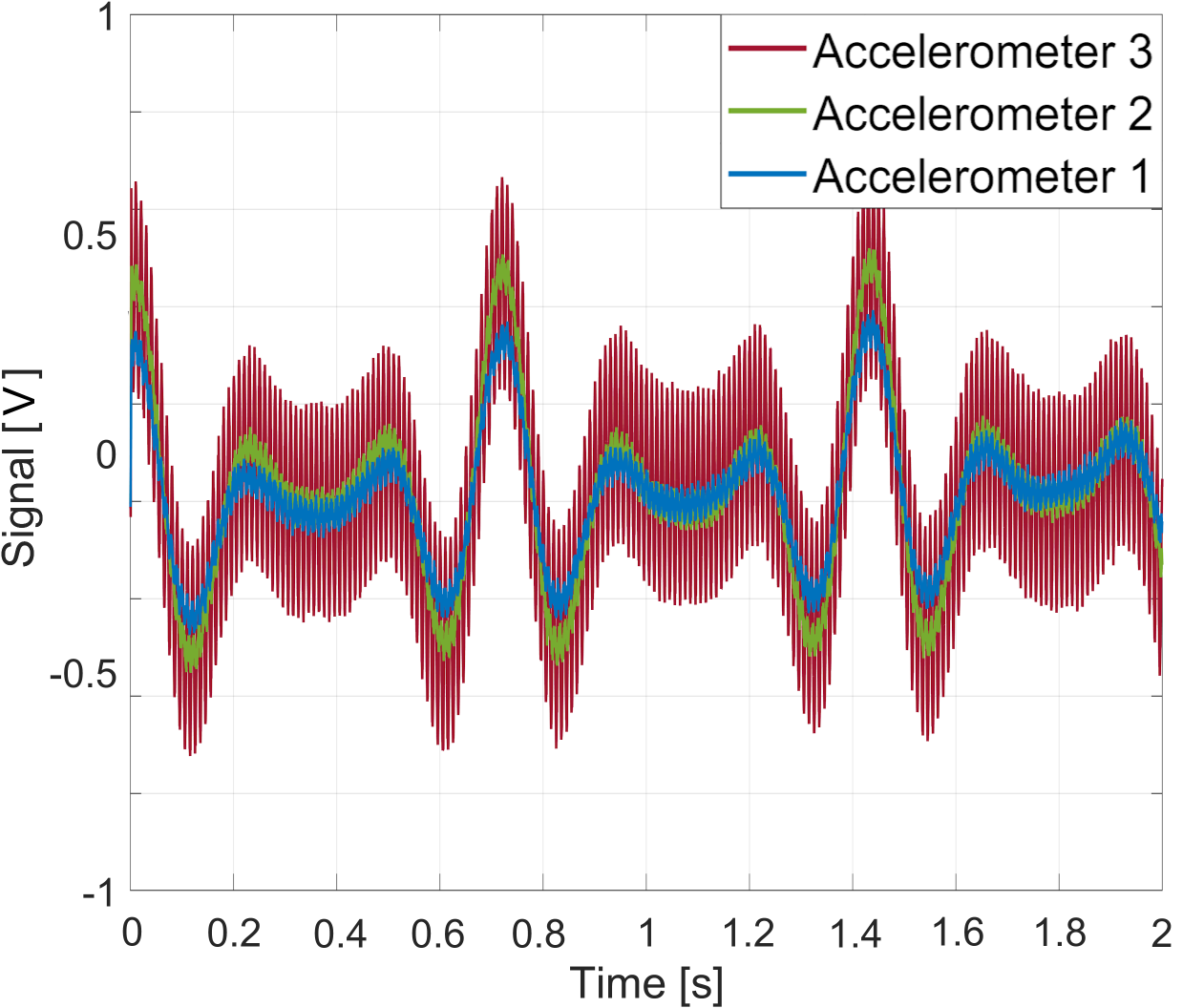}
\caption{}
\end{subfigure}
\begin{subfigure}{\linewidth}
 \includegraphics[width=\linewidth]{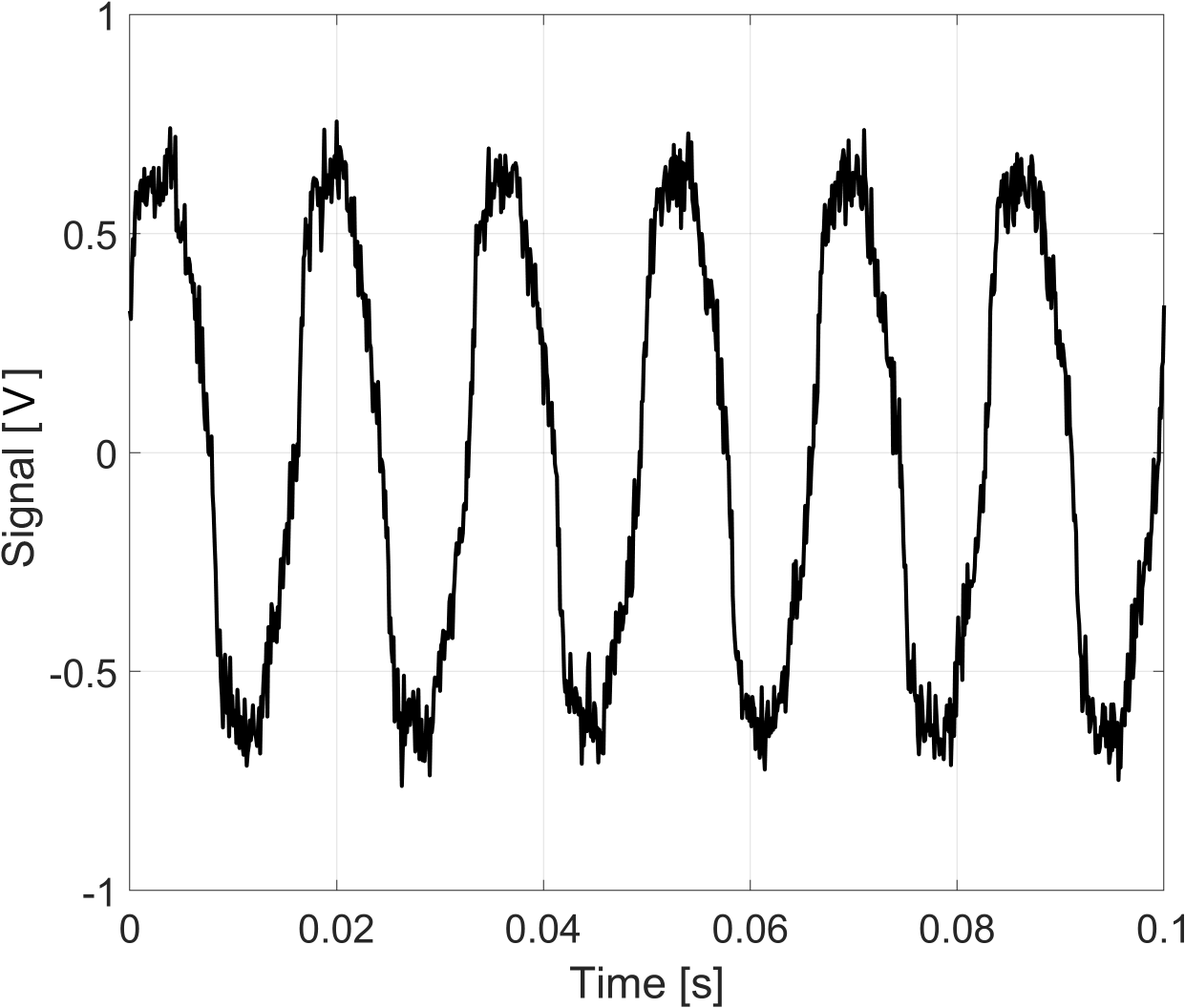}
\caption{}
\end{subfigure}
\caption{(a) Simulated accelerometer signals. These signals simulate vibrations in the laboratory from sources such as pulse tubes (PTs), fans, and transient events such as earthquakes. (b) Simulated antenna signal. The signal simulates electromagnetic pickup in the laboratory with a large 60 Hz component.}\label{fig:auxts}
\end{figure}

\begin{figure}[!ht]
 \centering
 \includegraphics[width=\linewidth]{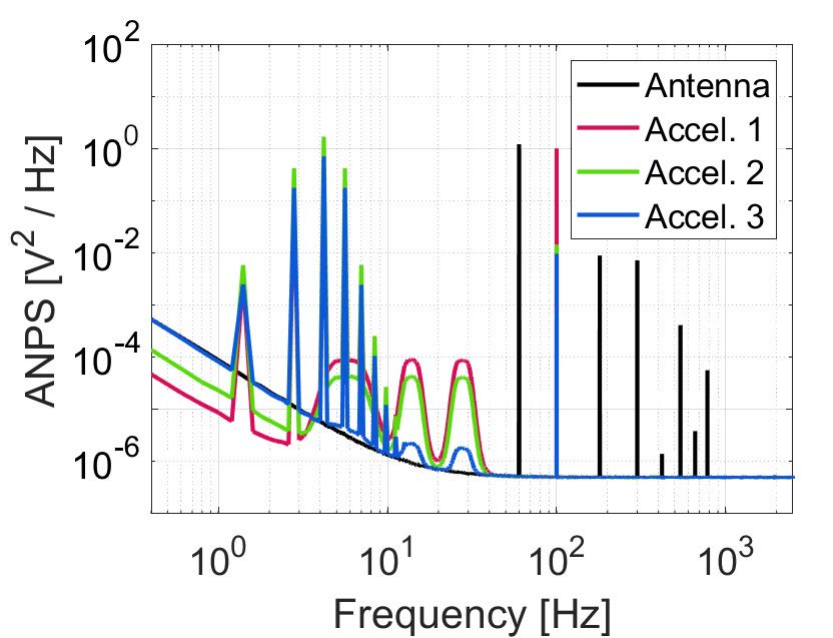}
 \caption{Average noise power spectra of simulated auxiliary input signals. The simulated antenna signal (black) measures electrical noise and contains harmonics of 60 Hz. The simulated accelerometer accelerometer signals (red, green, and blue) measure vibrational noise. The vibrational noise is composed of harmonics of 1.4 Hz which represents PT noise, and broader peaks below 100 Hz which represent the excitation of mechanical resonances in the system due to sudden disturbances. }\label{fig:sim_ANPS}
\end{figure}

To create our mock antenna signal, $x_0[t]$, we add $1/f$ noise and Gaussian white noise to the electrical source signal. The addition of white noise simulates thermal noise in the antenna, and the $1/f$ noise represents the flicker noise often present in electronic amplifiers. We generate three accelerometer timestreams by individually passing the vibration source signal through unique band-pass filters and adding $1/f$ noise and white noise to each output. The band-pass filters represent the response function of each accelerometer, which we slightly vary for each signal. The resulting signals $x_{1}[t],x_{2}[t],x_{3}[t]$, correspond to accelerometer signals in the $\hat{x}$, $\hat{y}$, and $\hat{z}$ directions. An example of each input signal and their power spectra are shown in figures \ref{fig:auxts} and \ref{fig:sim_ANPS}.

\begin{figure}[!ht]
\begin{subfigure}{\linewidth}
 \includegraphics[width=\linewidth]{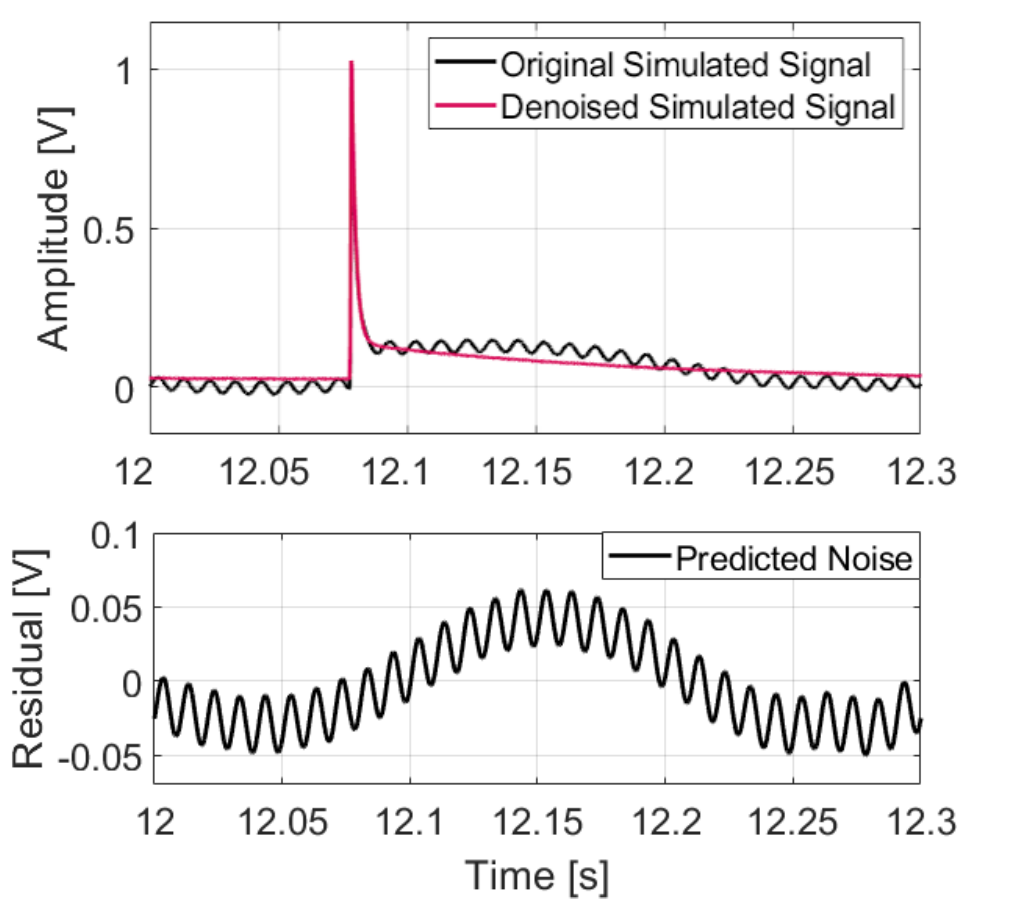}
 \caption{ }
\end{subfigure}
\begin{subfigure}{\linewidth}
 \includegraphics[width=\linewidth]{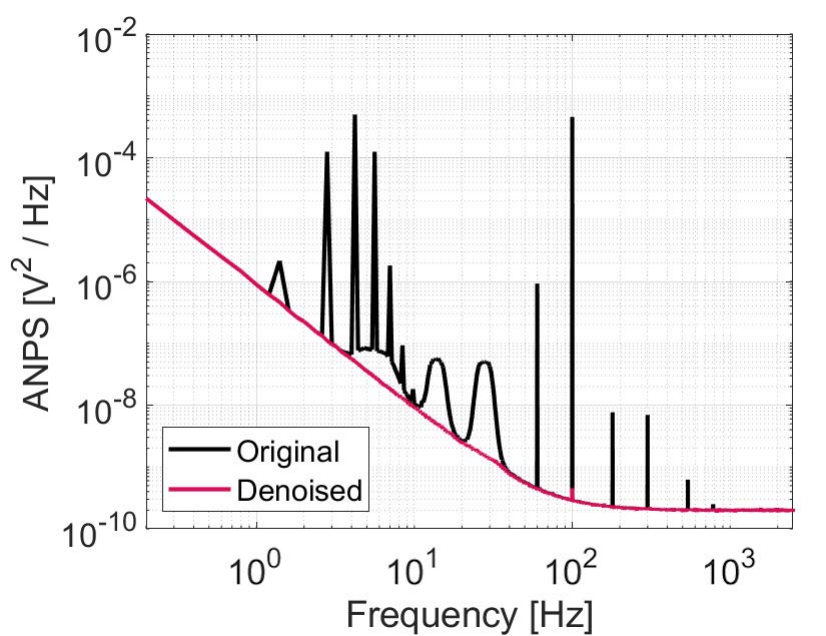}
 \caption{}
\end{subfigure}
\caption{(a) Simulated pulse event in the noisy detector signal (black) and denoised signal (red), and the noise predicted by the algorithm (below). (b) ANPS of the simulated detector signal before denoising (black) and after denoising (red). The noise that is correlated with the auxiliary devices is removed.}\label{fig:predicted}
\end{figure}

To create the mock detector signal, we again generate random white noise and $1/f$ noise. These terms are similar to some of the noise terms used in \cite{Cuore_Noise}. We then take the vibrational source signal and pass it through a unique band-pass filter. This models the detector's response to vibrations, which should be different from that of the auxiliaries. We add the result and the electrical source signal to the detector signal. Finally, we take a template pulse shape and generate a train of pulses which are randomly distributed in time with the caveat that the minimum time between pulses is longer than the length of the template to prevent pileup events. The pulse amplitude is randomly chosen to be 1 or 2 units with equal probability. The pulse amplitudes represent physical events with energies $E_0$ and $E_1 = 2E_0$ assuming linearity of the detector response. After generating the set of valid noise events and pulse events, we apply the noise decorrelation algorithm to the detector signal. Using the simulated antenna and accelerometer signals as auxiliary inputs, we construct the transfer functions $H_{x_iy}$ following equations \ref{eq:giy}, \ref{eq:gij}, and \ref{eq:hxy}, where the expectation values are taken over 5-second long noise windows. The results of the noise decorrelation on the output signal are shown in Figure \ref{fig:predicted}.

We now seek to show that the amplitude reconstruction and resolution of the simulated data is improved with the addition of the denoising algorithm. To perform our amplitude estimate, we use the detector noise power spectrum and a time-domain average of the pulses from the detector timestream to construct the matched filter, also called the optimal filter (OF) \cite{matched_filter, north, gatti, branca2020performance}. Applying the OF to the detector signal produces a new signal for which the signal to noise ratio is maximized \cite{OF}. Provided that the OF has unity gain, the maximum values of the filtered pulses are the optimal estimators of the pulse amplitudes \cite{golwala}. We can empirically measure the OF noise resolution by choosing an amplitude value from the same random time point within each noise window. The standard deviation of these values is a good approximation for the OF noise resolution of the detector. 

\begin{table*}
\begin{center}
\captionof{table}{\label{Tab:sim_results_table} Amplitude resolution and reconstruction error of noise events and simulated pulse events after denoising technique is applied. The reconstruction error is the difference between the measured pulse amplitude and the simulated pulse amplitude. The denoising algorithm reduces the magnitude of the reconstruction error and improves the resolution at all amplitudes. The resolutions at all amplitudes are compatible before and after denoising. This is expected because no energy-dependent resolution scaling is built into the simulation.}
\begin{tabular}{l | c c c }
\hline
 Simulated Pulse Amplitude [V] & 0  (Noise) & 1 & 2 \\ [0.5ex] 
 \hline
 Original Reconstruction Error $[\mu V]$ & $ 9 \pm 5$ & $-210 \pm 10$ & $-460 \pm 10$ \\
 Denoised Reconstruction Error $[\mu V]$ & $-7 \pm 4$ & $ -8 \pm 10$ & $  -40 \pm 10$ \\
 \hline
 Original Resolution $[\mu V]$ & $684 \pm 3$ & $680 \pm 10 $ & $680 \pm 10$ \\
 Denoised Resolution $[\mu V]$ & $527 \pm 3$ & $530 \pm 10 $ & $520 \pm 10$ \\
 \hline
\end{tabular}
\end{center}
\end{table*}

To analyze the results of the noise decorrelation algorithm, we run two analyses in parallel. First we construct the OF using the original simulated signal and use it to reconstruct the pulse amplitudes and OF noise resolution. Separately, we construct a new OF using the denoised data and apply it to the denoised data. The results before and after denoising are shown in Table \ref{Tab:sim_results_table}. One important ramification of the algorithm is that it can change the mean value of the OF amplitudes. This is expected because the denoising removes non-stationary noise from the system and because the average pulse created after the denoising is less correlated with the remaining noise. In principle, the denoised average pulse is a better representation of the physical pulse, and in practice the algorithm serves to unbias the data. The amplitude resolution is improved at the noise peak and both pulse amplitudes. In this particular iteration of the simulation, the amplitude resolution consistently improves by 22--23\%. Both before and after denoising, the amplitude resolutions across all three values are compatible with each other. This is also expected because no energy-dependent resolution scaling is built into the simulation. 

\section{Denoising Experimental Data}\label{sec:berkeley_denoising}

\subsection{Experimental Setup}\label{sec:berkeley_setup}

We performed the measurements to demonstrate the denoising algorithm in an Oxford Instruments Triton 400 dilution refrigerator cooled with a cryogen-free PT cooler at an operating temperature of $\sim$10 mK. Inside the refrigerator, we mounted a low-temperature light detector which serves as our detector of interest in this demonstration. The light detector is composed of a 45$\times$45$\times$0.5 mm$^3$ silicon wafer that absorbs the light increasing its temperature. This temperature change is measured by a 3$\times$3$\times$1 mm$^3$ NTD germanium thermistor which is epoxied to the wafer. The silicon wafer was rigidly connected to a square copper frame that was connected to the experimental stage of the dilution refrigerator. The front-end electronics used in this setup are identical to those used in CUORE \cite{cuore_frontend}. We ran a fiber-optic cable to near this detector with the axis of the cable approximately normal to the substrate surface. We connected the fiber to a commercially available LED with a wavelength of ($600 \pm 12$) nanometers to generate a train of LED light pulses with durations and separation times that can be easily controlled by a signal generator with pulse width modulation (PWM). This allowed us to create periodic light signals which we could use to characterize our detectors.

Alongside the light detector, we also mounted another NTD-Ge thermistor which was adhered to a small shard of a silicon wafer. This “NTD vibration sensor” was not rigidly connected to the copper frame, but instead left free to vibrate during the data-taking. Because the NTD vibration sensor was adhered to a relatively small wafer, we expected that the rate of pulses it observed would be negligible compared to that of the light detector. We could therefore use it as another auxiliary device while making cuts to the data in the rare case of a pulse event on the auxiliary device. This was a useful addition to the setup because we expected that the vibrations it measured would be closely correlated to those measured in the light detector due to their close proximity. We also expected that the electrical noise on each of the NTD signals will be similar since they are being read out in identical ways.

\begin{figure}[!ht]
\begin{subfigure}{\linewidth}
 \includegraphics[width=\linewidth]{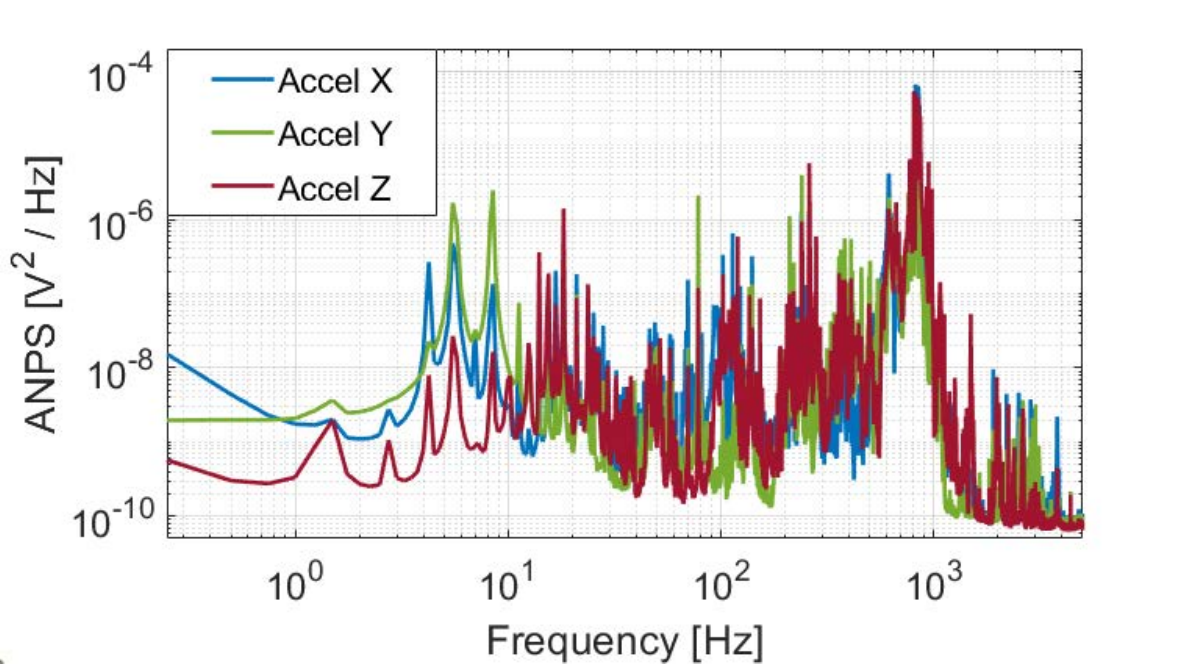}
 \caption {}
\end{subfigure}
\begin{subfigure}{\linewidth}
 \includegraphics[width=\linewidth]{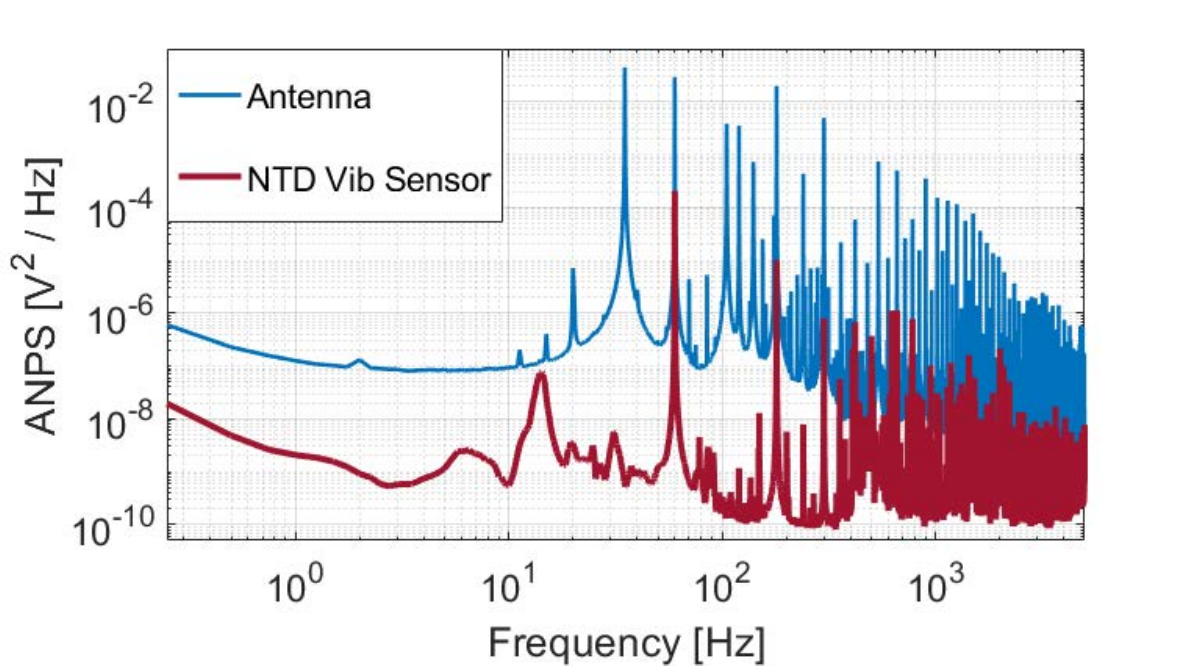}
 \caption {}
\end{subfigure}
\caption{(a) ANPS of accelerometers used in the experiment. PT noise at harmonics of 1.4 Hz and many other high-frequency peaks are visible. The noise power from $\sim$10 Hz to 100 Hz is reduced in the $\hat{z}$-oriented accelerometer because the experimental setup is equipped with dampeners that reduce vertical vibrations. (b) ANPS of antenna and NTD vibration sensor used in the experiment. The antenna measured a signal dominated by components at 35 Hz, 60 Hz, and their harmonics. The NTD vibration sensor signal is dominated by the odd harmonics of 60 Hz, though there is also a source of broadband noise between 10 and 20 Hz.}\label{fig:aux_berk}
\end{figure}

\begin{figure}[!ht]
\begin{subfigure}{\linewidth}%
 \includegraphics[width=\linewidth]{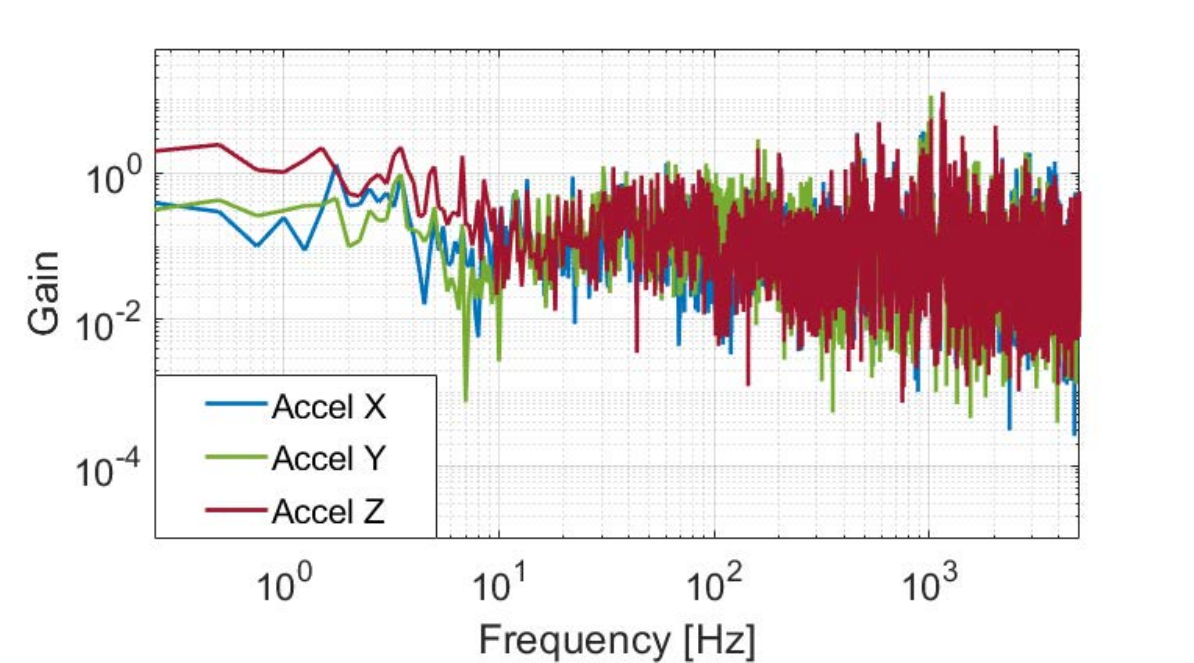}
 \caption {}
\end{subfigure}
\begin{subfigure}{\linewidth}%
 \includegraphics[width=\linewidth]{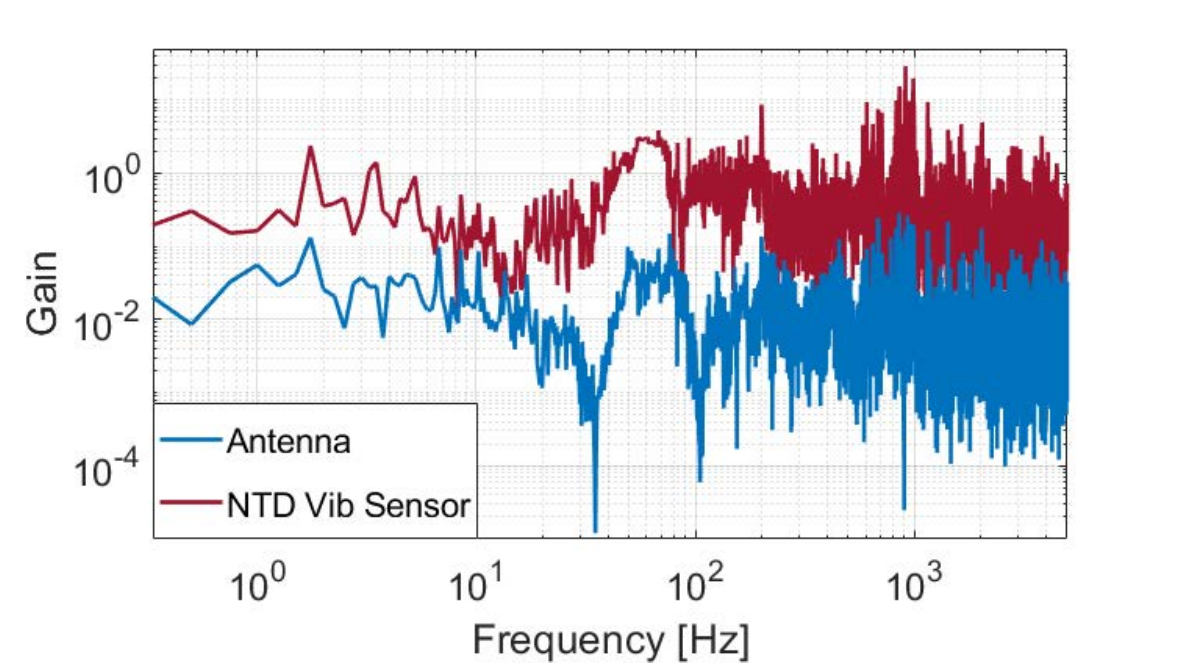}
 \caption{}
\end{subfigure}
\caption{(a) Accelerometer transfer functions with light detector. The peaks at the harmonics of 1.4 Hz, particularly in the $\hat{z}$-oriented accelerometer, indicate a correlation between the PT noise observed in the accelerometers and the microphonic noise in the light detector. (b) Antenna and NTD vibration sensor transfer functions with light detector. Note the wide-band region of high gain near 60 Hz for both of these devices. The large troughs at 35 Hz and 105 Hz indicate that the antenna signal is not correlated with the light detector at these frequencies.}\label{fig:TF_berk}
\end{figure}

To measure the electrical noise present in our system, we attached an antenna to our DAQ and placed it inside the Faraday cage which houses the experiment. The antenna we used was a hand-made Helmholtz coil with $\sim 50$ turns and a radius of $\sim 30$ cm, and we amplified the antenna signal by a factor of 5,000. To measure the vibrational noise, we used three PCB 393B31 accelerometers \cite{PCBaccels} mounted triaxially to the 300 K plate of the dilution refrigerator. The accelerometer signals were amplified with a PCB 482C15 signal conditioning pre-amplifier \cite{PCBamp} and read out to the same DAQ system as the antenna, NTD vibration sensor, and light detector. The ANPS of the auxiliary devices are shown in Figure \ref{fig:aux_berk}. 

As a starting ground for testing the algorithm, we collected data while a ground loop persisted in our readout chain. This gave rise to noise peaks at 60 Hz and subsequent harmonics from the AC power line with sufficient power to dominate the detector signal. In addition to the electrical noise, we observe noise peaks at 1.4 Hz and subsequent harmonics due to vibrations from the PT which vibrate the internal cryogenic structure, including the detector and the cold electronics. To construct an average noise template, we first took a long run of data without injecting any pulses to ensure a sufficient number of noise events free of pulses. The data from these runs were used to construct the transfer functions, $H_{x_iy}$, which are shown in Figure \ref{fig:TF_berk}. After, we acquired two runs of data, each with a set of LED pulses at a fixed frequency (1.7 Hz). The first set of LED pulses had a duration of 100 $\mu$s, and the second set had a duration of 200 $\mu$s and thus twice the energy of the first set. We split the PWM signal and acquired it with our DAQ to efficiently trigger on the light pulses with few backgrounds. We denoised these LED runs by applying the previously calculated transfer functions to the auxiliary data from the LED runs and subtracting the result from the light detector signal.

\subsection{Analysis Techniques}\label{sec:berk_techniques}

Since we were able to trigger the LED pulses, we could still analyze our signal despite the low SNR in the system. We used the same analysis procedure described in Section \ref{sec:sim}, but this time we conducted three separate analyses. We again analyze the original data and denoised data, but we also analyze the data after applying a notch filter centered at 60 Hz to the signal. Since the signal is dominated by 60 Hz noise, one might expect that notch filtering the data will lead to an improvement comparable to that of the noise decorrelation algorithm. We selected 0.5-second windows between LED pulses and making RMS-based cuts to veto windows with spurious events (e.g. caused by cosmic rays) either on the detector signal or NTD vibration sensor. We then construct the average noise power spectrum for each of the three signals. We use the LED triggers to build an average pulse which is also 0.5 seconds long. Together, the noise events and average pulse template allow us to denoise the data and build the OF. We separately construct the OF using the original data, the band-pass filtered data, and the denoised data.

To align the gains of the three filters, we perform a pseudo-calibration assuming that the second set of LED pulses has twice the energy of the first one. We align the LED pulse amplitudes to arbitrary energy values of $E_0$ and $2E_0$. We construct histograms of the amplitudes obtained by each analysis of each LED event, then fit the amplitudes to a normal distribution using an unbinned fit. We then use a second-order polynomial fixed at the origin to align the measured amplitude values to a common scale. Following the work of \cite{singh2022}, one can model the energy resolution of the detector as $\sigma(E) = \sqrt{\sigma_0^2 + f(E)\sigma_E^2}$, where $\sigma_0$ is the baseline resolution of the detector, $\sigma_E$ is a scale factor that arises due to the stochastic nature of the LED signals, and $f(E)$ is a monotonically increasing function. We expect this behavior of the detector response because the magnitude of the Poisson fluctuations in the number of photons captured by the light detector will increase as the amplitude of the LED pulse increases. With this model, we expect that the denoising algorithm does not affect $f(E)$ or $\sigma_E$, but does lower the value of $\sigma_0$. We thus expect to see the greatest relative improvement in energy resolution at the noise peak where the energy-dependent effects are small. We define the ratio of resolutions $R(E) = \hat{\sigma}_{denoised}(E) / \hat{\sigma}_{original}(E)$ and calculate it at each energy for the three signal processing techniques.

\subsection{Results of Denoising}\label{sec:berk_results}

\begin{figure}[!ht]
 \centering
 \includegraphics[width=\linewidth]{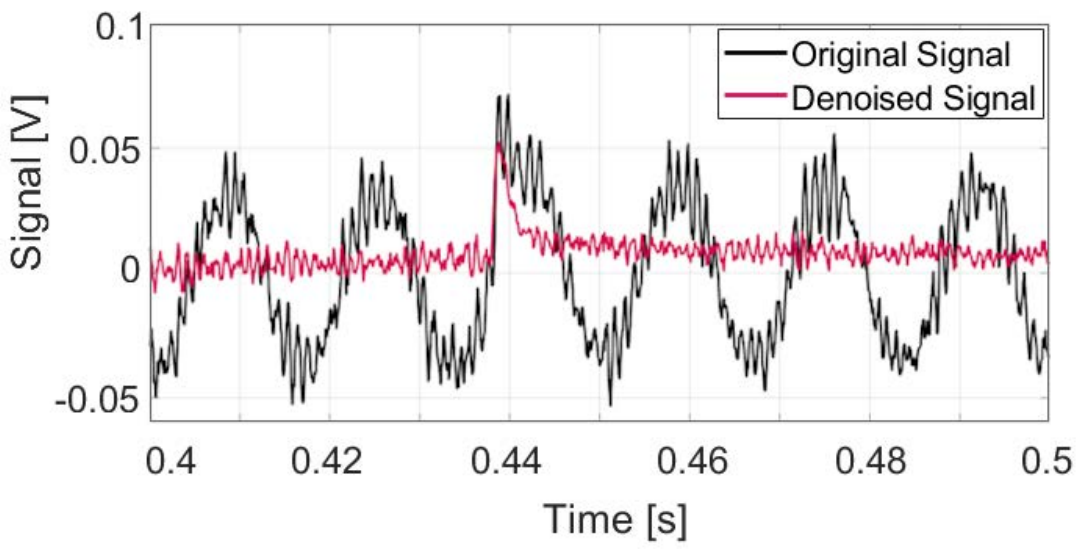}
 \caption{Light detector signal containing an LED pulse before and after denoising. Before denoising, the pulse is almost completely washed out by the noise which is dominated by a 60 Hz component. After denoising, the pulse is visible by eye.} \label{fig:berk_data_timestream}
\end{figure}

\begin{figure}[!ht]
\begin{subfigure}{\linewidth}
 \includegraphics[width=\linewidth]{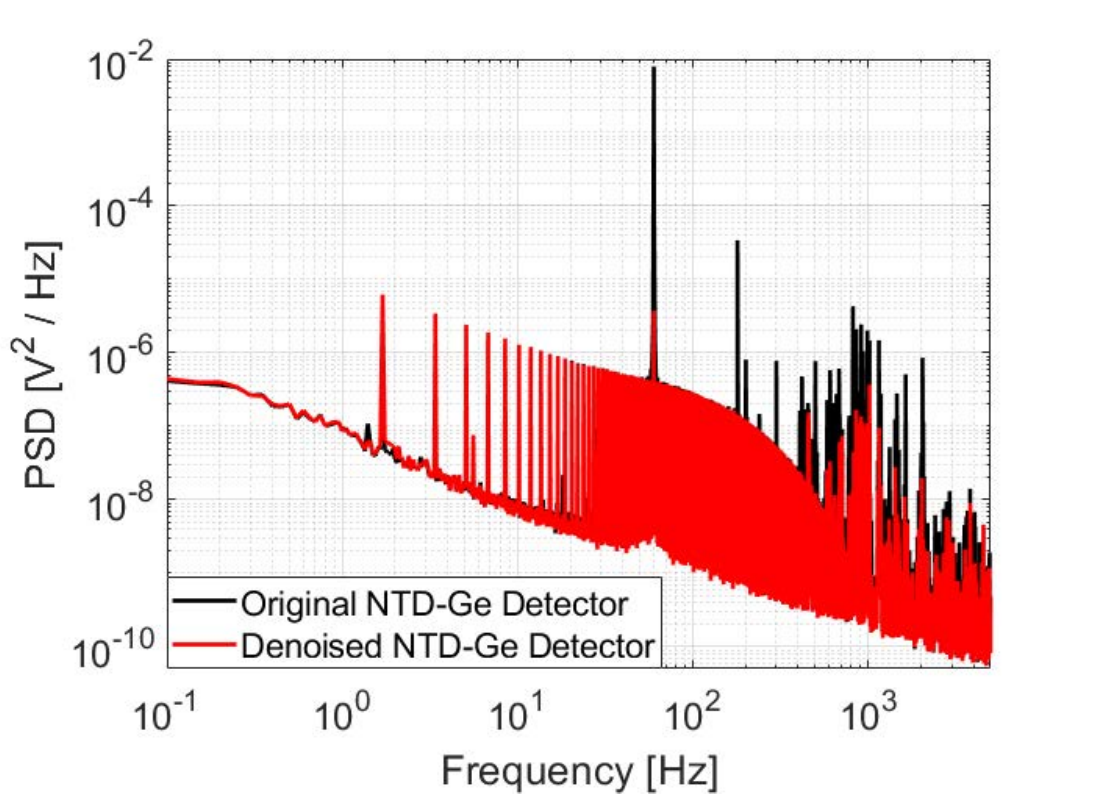}
 \caption{}
\end{subfigure}
\begin{subfigure}{\linewidth}
 \includegraphics[width=\linewidth]{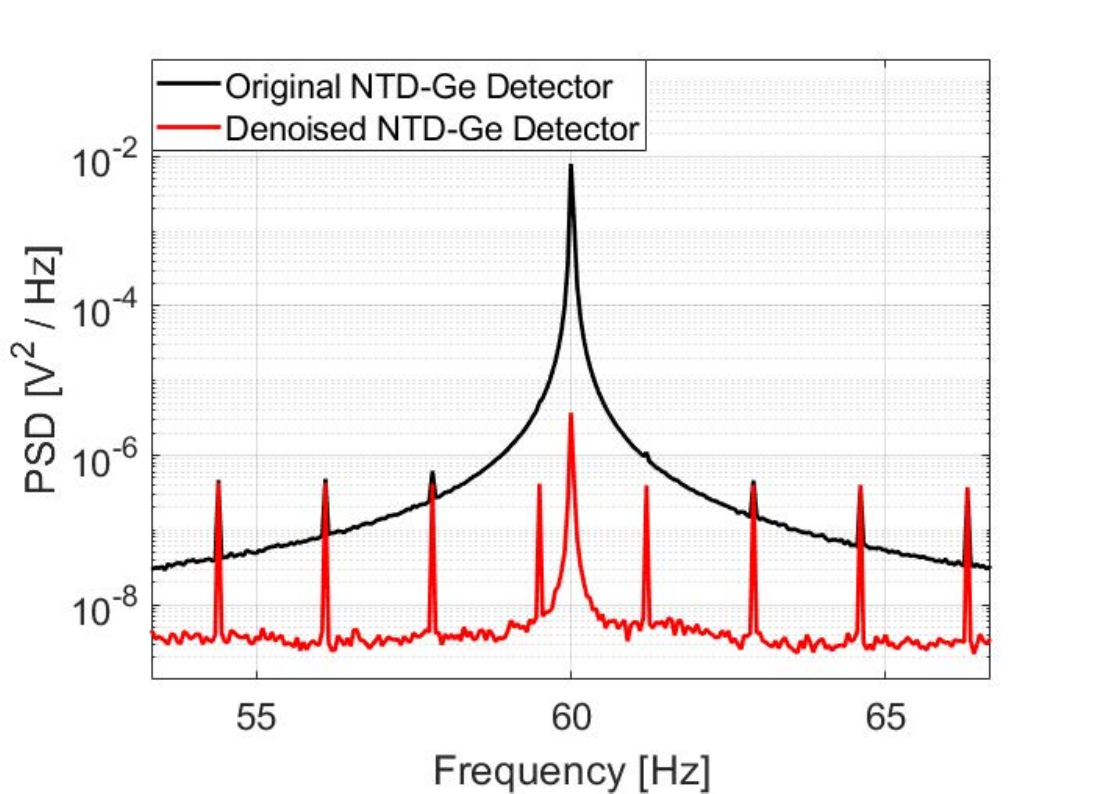}
\caption{}
\end{subfigure}
\caption{ (a) Average power spectrum of the light detector signal before and after denoising. The events used to build these power spectra include the data from the train of LED pulses which produces the evenly spaced peaks separated by 1.7 Hz. Noise peaks across the frequency spectrum, including the 1.4 Hz vibrational peak and the 60 Hz electrical peak, are eliminated after denoising. (b) Same power spectrum zoomed to the 60 Hz noise peak. The peak is reduced by $\sim$33 dB and the FWHM of the peak is reduced by approximately 10\%. The evenly spaced peaks due to the LED pulse train become more apparent after denoising. This structure is largely unaffected by the denoising, indicating an improved signal to noise ratio near 60 Hz.}
\label{fig:comp_berk}
\end{figure}

\begin{figure}[!ht]
 \centering
 \includegraphics[width=\linewidth]{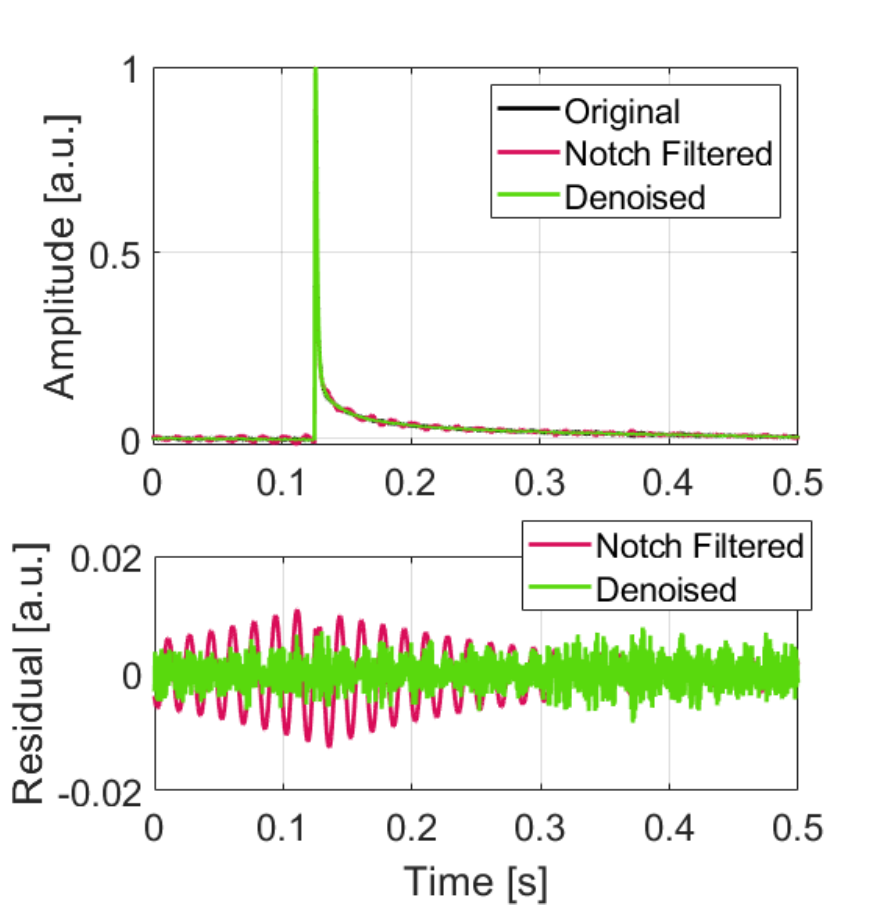}
 \caption{Average pulses constructed with the original, notch filtered, and denoised LED pulses on the light detector. The notch filter produces a ringing effect which can be seen in the plot of the residuals. The denoising procedure removes 60 Hz noise as well as other high-frequency noise from the average pulse, producing a smoother and more accurate pulse template for later use in the OF.} \label{fig:berk_AP}
\end{figure}

After denoising, there was a visible improvement in the signal-to-noise ratio of the detector as seen in Figure \ref{fig:berk_data_timestream}. We also observed a reduction in both vibrational and electrical noise in the light detector power spectrum (see Figure \ref{fig:comp_berk}). The vibrational peak at 1.4 Hz was eliminated, and many of the electrical peaks ranging from 60 Hz to 5 kHz were significantly reduced. The 60 Hz noise peak was reduced by 33 dB, and the full width at half-max (FWHM) was reduced by approximately 10\%. This peak was extremely wide due to small modulations of the ground loop frequency, but these modulations were measured by the antenna and the NTD vibration sensor so we could successfully remove them from the data. This is reflected in the wide-band region of high gain near 60 Hz in the transfer functions for both of these devices as was shown in Figure \ref{fig:TF_berk}. Outside of this region, the main effect of the denoising is a reduction in noise peaks. The algorithm does not have an apparent effect on the broadband detector noise. Before denoising, high frequency noise with a large 60 Hz component contaminates the average pulse. Applying the notch filter induces a ringing in all pulses which coherently add together when constructing the average pulse. When the denoising algorithm is applied, the noise in the average pulse is reduced without inducing any ringing (see Figure \ref{fig:berk_AP}).

We next examine the effect of the denoising algorithm on the resolution of the noise and LED pulse events. As expected, the resolution increases monotonically with energy. At all amplitudes, the denoising improves the measured detector resolution (see Table \ref{Tab:berk_results_table}). We see the greatest relative improvement at lower pulse amplitudes, which was also expected. It should be noted that the improvement in amplitude resolution at the first LED peak $(E = E_0)$ is compatible with the improvement at the noise peak, though it has a greater uncertainty.

\begin{figure}[!ht]
\centering
 \includegraphics[width=\linewidth]{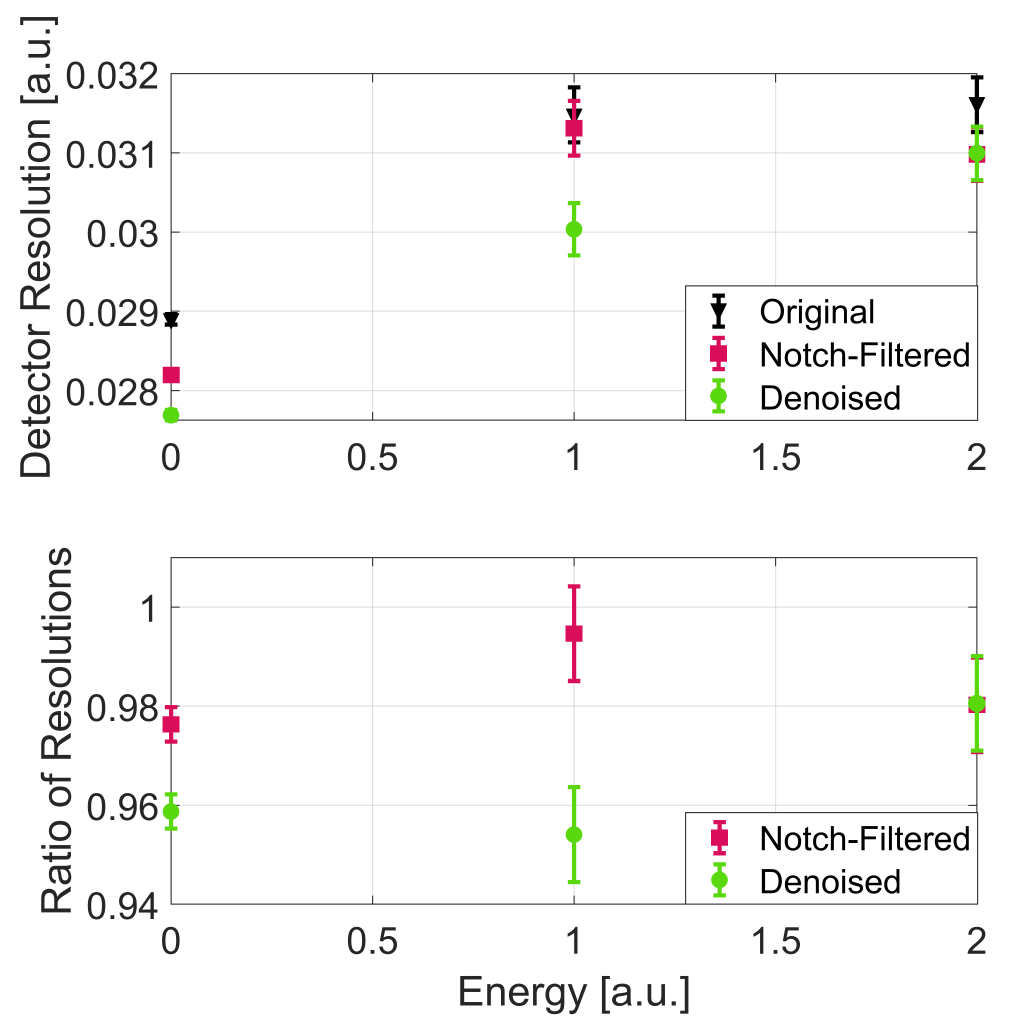}
\caption{Energy resolution of noise events and LED events before and after denoising or notch filtering is applied. The fact that the energy resolution increases with energy is expected due to the Poisson fluctuation of detected photons. At all energies, the denoising algorithm improves the resolution of the detector. At all but the highest energies, the algorithm out-performs the 60 Hz notch filtering of the data.
} \label{fig:berk_energy_res}
\end{figure}

We also observe that the denoising algorithm performs as well as or better than the notch filtering approach (see Figure \ref{fig:berk_energy_res}). This effect is especially visible at the noise peak, where the improvement in the detector resolution from the denoising (4.1\%) is 1.7 times that of from the notch filtering (2.4\%). At the highest-amplitude peak, we do not expect the denoising to have as much of an effect on the resolution, but we still observe an improvement. At this amplitude, we see that the denoising and notch filtering have similar effects on the data These results are summarized in Table \ref{Tab:berk_ratio_table}.

\begin{table*}
\begin{center}
\captionof{table}{\label{Tab:berk_results_table} Best-fit estimators of energy resolution with and without denoising applied. Here $\hat{\sigma}_{0}$ is the best estimator of the energy resolution of the detector noise events. All energies are scaled to that of the first LED pulse amplitude. All energy resolutions are calculated from the optimally filtered data.}
\begin{tabular}{l | c c c }
\hline
 Signal Type & $\hat{\sigma}_0(E = 0)$ & $\hat{\sigma}(E = E_0)$ & $\hat{\sigma}(E = 2E_0)$ \\ [0.5ex] 
 \hline
 Original       & $(2.89 \pm .01)\times 10^{-2}$ & $(3.15 \pm .04)\times 10^{-2}$ & $(3.16 \pm .04)\times 10^{-2}$ \\
 Notch Filtered & $(2.82 \pm .01)\times 10^{-2}$ & $(3.13 \pm .03)\times 10^{-2}$ & $(3.10 \pm .03)\times 10^{-2}$ \\
 Denoised       & $(2.77 \pm .01)\times 10^{-2}$ & $(3.00 \pm .03)\times 10^{-2}$ & $(3.10 \pm .03)\times 10^{-2}$ \\
 \hline
\end{tabular}
\end{center}
\end{table*}

\begin{table*}
\begin{center}
\captionof{table}{\label{Tab:berk_ratio_table} Ratio of denoised and notch filtered energy resolutions to original energy resolution. The denoising algorithm outperforms the notch filter given its capability to remove noise across a large range of the frequency domain. Note that the main effect of the denoising is in the reduction in baseline resolution $R_0$. This is expected given that the resolution is dependent on the LED amplitude due to the Poisson fluctuation of detected photons.}
\begin{tabular}{l | c c c }
\hline
 Signal Type & $\hat{R}_{0}$ & $\hat{R}(E = E_0)$ & $\hat{R}(E = 2E_0)$ \\ [0.5ex] 
 \hline
 Notch Filtered & 0.976 $\pm$ 0.003 & 0.99 $\pm$ 0.01 & 0.98 $\pm$ 0.01 \\
 Denoised       & 0.959 $\pm$ 0.003 & 0.95 $\pm$ 0.01 & 0.98 $\pm$ 0.01 \\
 \hline
\end{tabular}
\end{center}
\end{table*}

It is important to note that the improvement in resolution is $\leq 5\%$ in all cases. This is also expected since the OF is by definition the filter resulting in the highest SNR given a particular choice of noise and pulse template. The improvements from the denoising are thus limited by the amount of noise that can be removed from the noise and signal templates, which can in turn be limited by the "SNR" of the auxiliary devices, i.e. the ratio of the amount of noise correlated with the detector of interest to the amount of uncorrelated noise. Having verified the efficacy of the denoising technique on a controlled experiment, we applied the method to data from the CUORE experiment.

\section{Denoising CUORE Data}\label{sec:cuore_denoising}

\subsection{The CUORE Experiment}\label{sec:cuore}

CUORE (Cryogenic Underground Observatory for Rare Events) \cite{cuore2022search} is an ongoing experiment at the Gran Sasso National Laboratory (LNGS) primarily searching for the lepton number violating process of neutrinoless double beta decay $(0\nu\beta\beta)$ \cite{elliott2002double, dolinski2019neutrinoless}. The discovery of such a process would be a clear experimental signature of physics beyond the Standard Model and may provide insight into the nature of the absolute neutrino mass scale \cite{schechter1982neutrinoless, vergados2016neutrinoless, 0nuphysics1, 0nuphysics2}. The sensitivity to such a discovery requires detectors that have very good energy resolution. For example, in the case of a finite, non-zero number of expected background counts, the sensitivity to the half-life of $0\nu\beta\beta$ scales as $(\Delta E)^{-\frac{1}{2}}$ \cite{nutini2018performance}, where $\Delta E$ is the energy resolution of the system at the expected peak energy $Q_{\beta\beta}$. 

CUORE employs cryogenic calorimeters consisting of a 5x5x5 cm$^3$ TeO$_2$ crystal, a silicon resistive heater, and an NTD-Ge thermistor that measures the small temperature rise in the crystal when energy from a nuclear decay or particle interaction is deposited in it. Each of these calorimeter signals is then read out as an individual channel and digitized with a sampling frequency of 1 kHz. As mentioned before, CUORE uses four Cryomech PT 415 cryocoolers with fundamental frequencies of 1.4 Hz to cool the cryostat. The vibrations induced by the PTs are a large source of noise due to CUORE's relatively small bandwidth ($\sim$80 Hz), and active steps are already taken to reduce this PT noise during data-taking \cite{cuorePTnoise}. Nonetheless, this noise persists in many channels in the CUORE data. 

To measure sources of microphonic noise, we installed three triaxially-oriented accelerometers identical to the ones described in Section \ref{sec:berkeley_setup} on the structural support frame of the cryostat and read the three signals out to the CUORE DAQ system. We also installed a triaxial Sara Electronics SS-10 seismometer \cite{seismoSource} to the CUORE suspension system and read it out to the same DAQ. We use these devices as the six input signals when we apply the noise decorrelation algorithm. We also motivate the use of a nonlinear version of the algorithm for thermal detectors. Using 24 hours of data from one CUORE detector channel, we demonstrate both the linear and nonlinear versions of the denoising algorithm using the accelerometer and seismometer signals as inputs. We then analyze the noise events and so-called ``pulser'' events to evaluate the performance of the algorithm.

\subsection{CUORE Pulser Events}\label{sec:pulsers}

To account for changes in the detector response due to thermal fluctuations, short pulses of known voltages are applied to the CUORE detector at regular intervals using the silicon heaters mentioned in Section \ref{sec:cuore}. This results in \textit{pulser} events in the CUORE channels. Since the injection method is practically identical for each pulser event, we expect their resolutions to be comparable to the noise resolution of the channel, so they make good candidates for verifying the performance of the denoising algorithm. The particular channel we choose has pulser amplitudes during this run. The two amplitudes are comparable to physical pulses with energy depositions of 960 keV and 3500 keV. These energy values were determined during the online CUORE data analysis. We will use the two pulser event energies and the noise peak to calibrate the detector. 

\subsection{The Nonlinear Algorithm}\label{sec:nonlin}

It is expected that quadratic mixing of acceleration signals should occur in the CUORE thermal detectors. This is motivated both by the non-zero bicoherence observed in CUORE (see Appendix \ref{app:bicoherence}) and also from first principles. A simplified explanation for this is that any power incident on the detector due to vibrations heats a thermal detector regardless of directionality of the vibration. This power is non-trivially related to the square of the waveform describing this vibration (see Appendix \ref{app:mix} for a detailed explanation). Here we use the square of each auxiliary signal $x_i^2[t]$ as a proxy for the thermal response to vibrational noise. Approaches similar to this which analyze nonlinear systems by including higher-order powers of the input signals are well described in \cite{rice1988generalised, bendat1982spectral}. We still keep the original auxiliary signals $x_i[t]$ since they may be separately correlated with the detector noise, for example due to capacitive pickup in the detector wiring. We proceed with the denoising algorithm using the new set of input signals, which is twice as large as the original set. In practice the Fourier amplitudes $X_i[f]$ and $X_i^2[f]$ can be strongly correlated, so one must ensure that the matrices are well-conditioned before proceeding. While we do not expect this model to capture the full nonlinear behavior of the thermal response to the input signals, we do expect to see an improvement in the noise decorrelation over the linear model.

\subsection{Results of Denoising}\label{sec:cuore_results}

\begin{figure}[!ht]
\begin{subfigure}{\linewidth}
 \includegraphics[width=\linewidth]{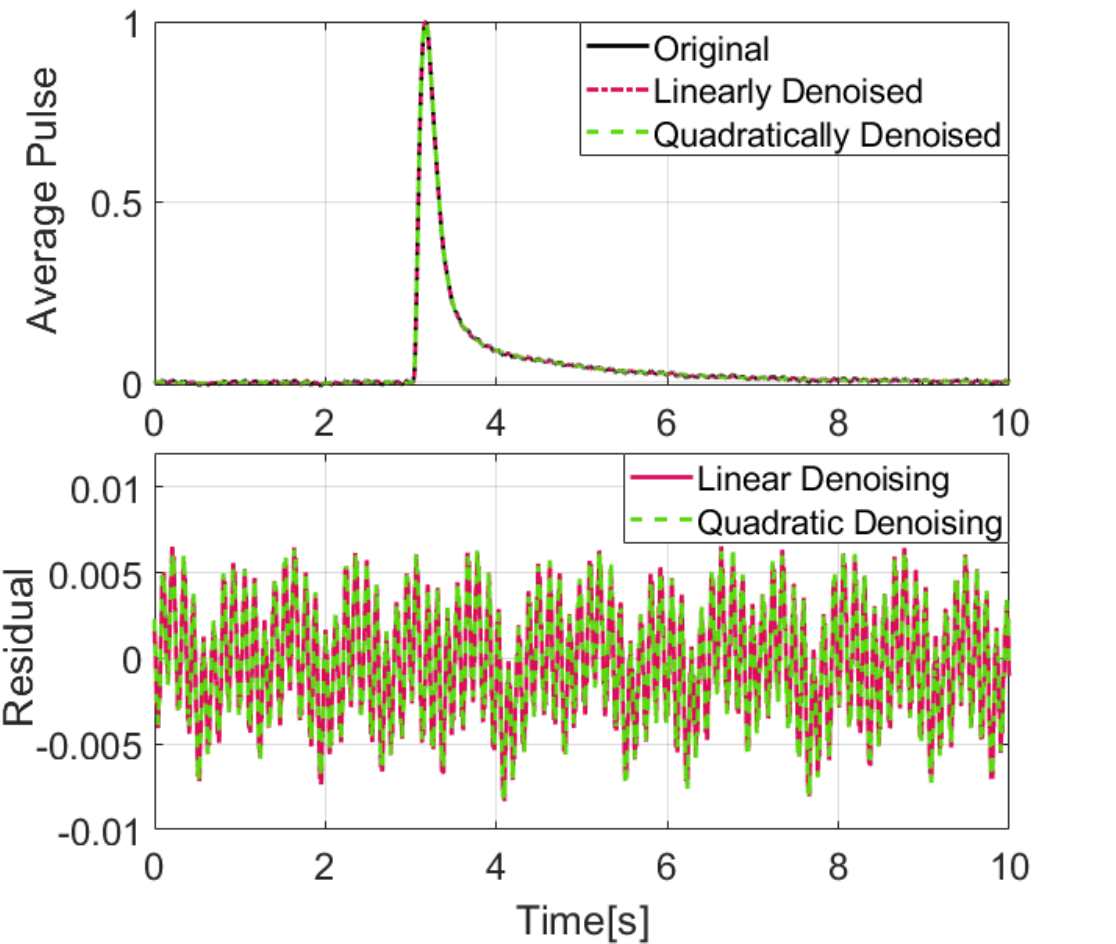}
 \caption{}
\end{subfigure}
\begin{subfigure}{\linewidth}
 \includegraphics[width=\linewidth]{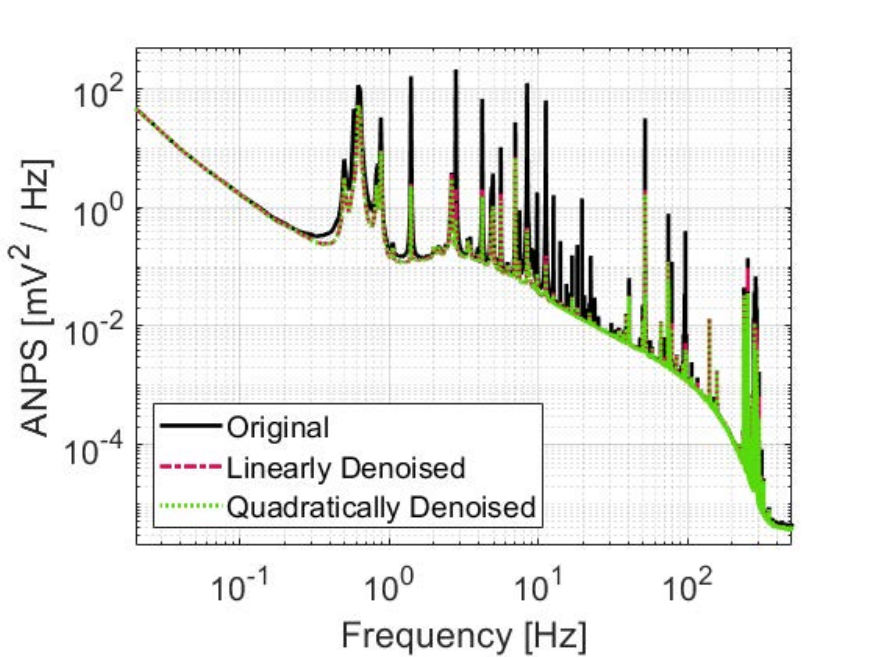}
 \caption{}
\end{subfigure}
\caption{Effects of linear and nonlinear denoising techniques before constructing the OF. (a) Average pulse created using the original, linearly denoised, and nonlinearly denoised data. The denoising procedure removes noise containing many harmonics of 1.4 Hz from the average pulse. (b) ANPS of a CUORE channel before denoising and after the two versions of the denoising algorithm. The linear version removes noise across the signal band, from less than 1 Hz to several hundred Hz. At several frequencies including 2.8 Hz, the nonlinear version of the algorithm further reduces the noise.}\label{fig:cuore_AP_ANPS}
\end{figure}

\begin{figure}[!ht]
\begin{subfigure}{\linewidth}
 \includegraphics[width=\linewidth]{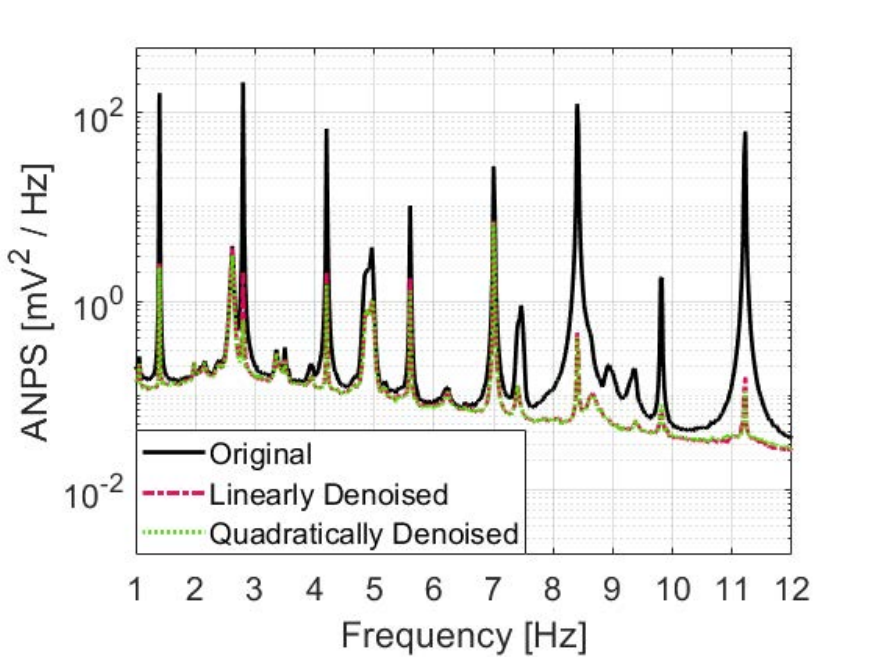}
 \caption{}
\end{subfigure}
\begin{subfigure}{\linewidth}%
 \includegraphics[width=\linewidth]{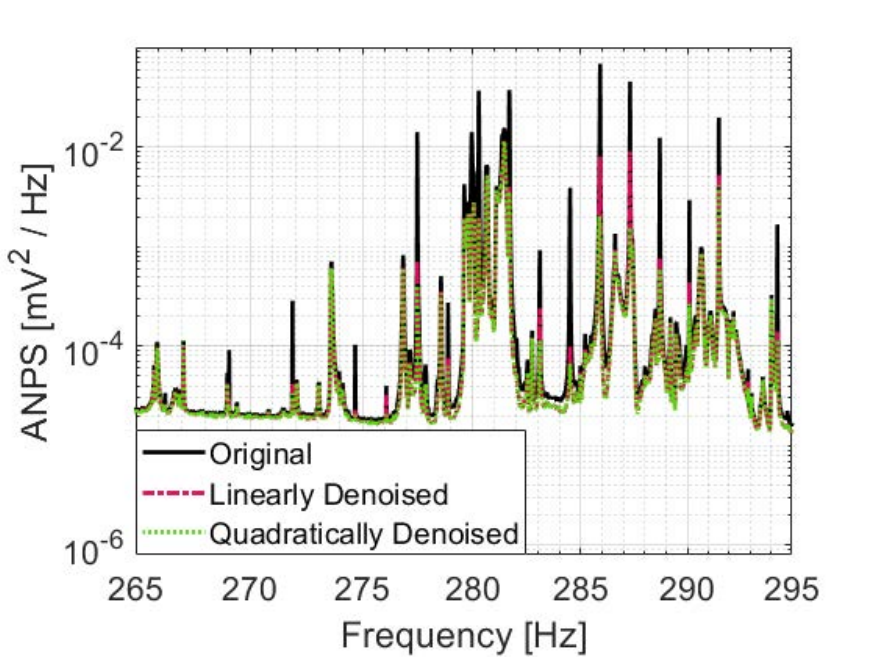}
 \caption{}
\end{subfigure}
\caption{(a) ANPS of a CUORE channel in the region dominated by PT noise at the harmonics of 1.4 Hz. The nonlinear version of the algorithm reduces the noise peaks only slightly more than the linear version. This difference is most noticeable at 2.8, 4.2, and 5.6 Hz. (b) The same CUORE ANPS around 280 Hz. Here, there are many peaks where the noise is reduced noticeably more by the nonlinear algorithm than by the linear one.}\label{fig:anps_zooms}
\end{figure}

The CUORE data have a much higher SNR than the data used to demonstrate the algorithm in Section \ref{sec:berkeley_denoising}. We are therefore able to trigger on pulse events directly and create a set of noise events. The only possible contamination of this set is from very low-amplitude pulses not detected by the CUORE triggering algorithm. This effect is expected to be small because the pulse rate is small ($\sim$3 mHz) and the small amount of signal power due to these pulses is negligible. We first run the algorithm with 50-second long noise windows using the accelerometer and seismometer signals as inputs, then we repeat the algorithm on the original data using the same signals and their squares as inputs. We conduct an independent analysis after each of the different denoising methods is applied to the raw data. We show that the denoising produces a cleaner pulse template as well as a reduced noise power spectrum. Comparing the noise power spectra, we see that the denoising performs excellently on the PT noise, but does not substantially impact the noise below 1 Hz and does little to change the broadband continuum noise below the various peaks (see Figure \ref{fig:cuore_AP_ANPS}). The nonlinear method is more effective than the linear one, though this is limited to specific bands in frequency space. The most noticeable differences between the linear and nonlinear algorithms are seen at the harmonics of 1.4 Hz and in a region around 280 Hz with many noise peaks (see Figure \ref{fig:anps_zooms}). This suggests the presence of a nonlinearity in the system from the auxiliary devices to the CUORE channel. This result is discussed further in Appendix \ref{app:bicoherence}.

\begin{figure}[!ht]
\centering
\includegraphics[width=\linewidth]{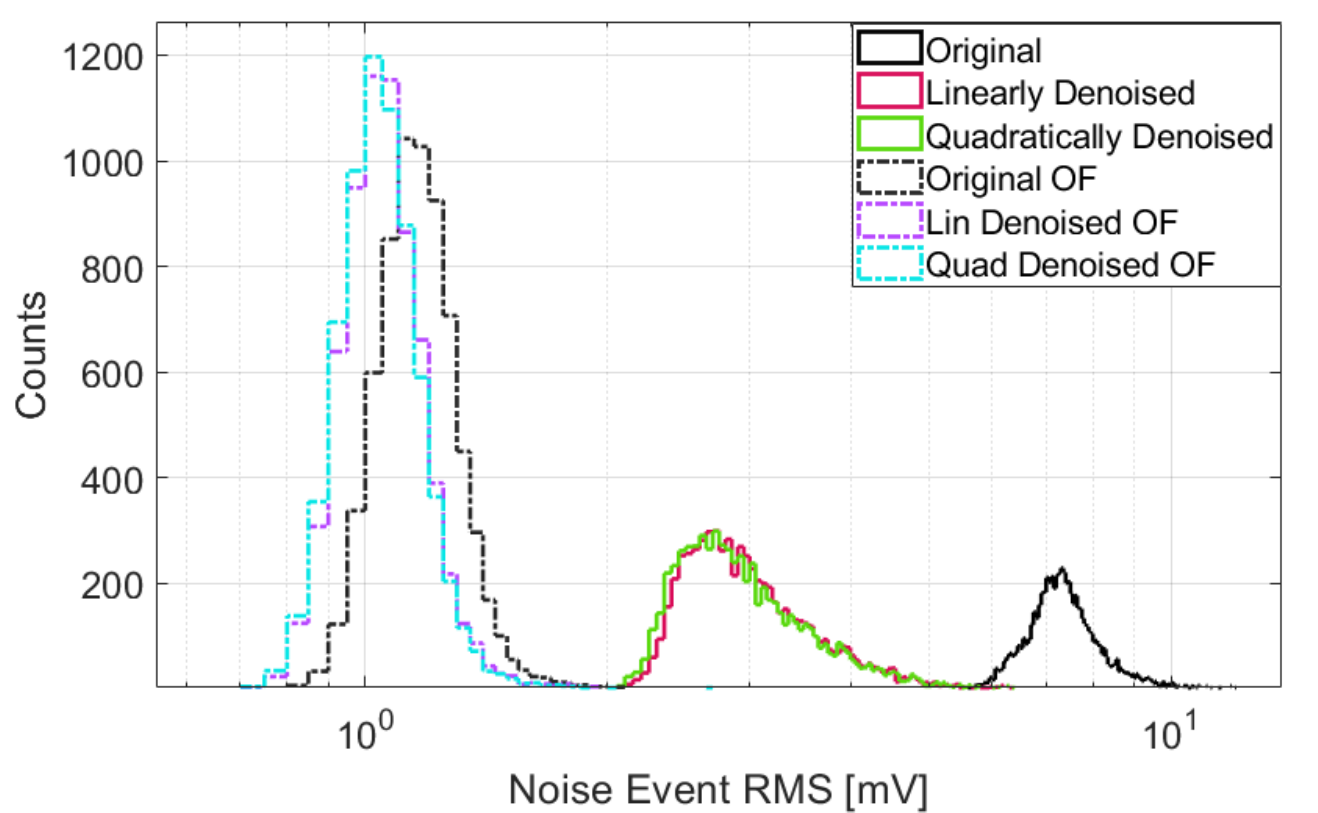}
\caption{Distribution of RMS of noise events before any signal processing, after denoising, and after applying the OF. The lowest noise RMS is achieved when the data are denoised with the nonlinear version of the denoising algorithm then optimally filtered.} \label{fig:cuore_allNoiseRMS}
\end{figure} 

\begin{figure}[!ht]
\centering
\includegraphics[width=\linewidth]{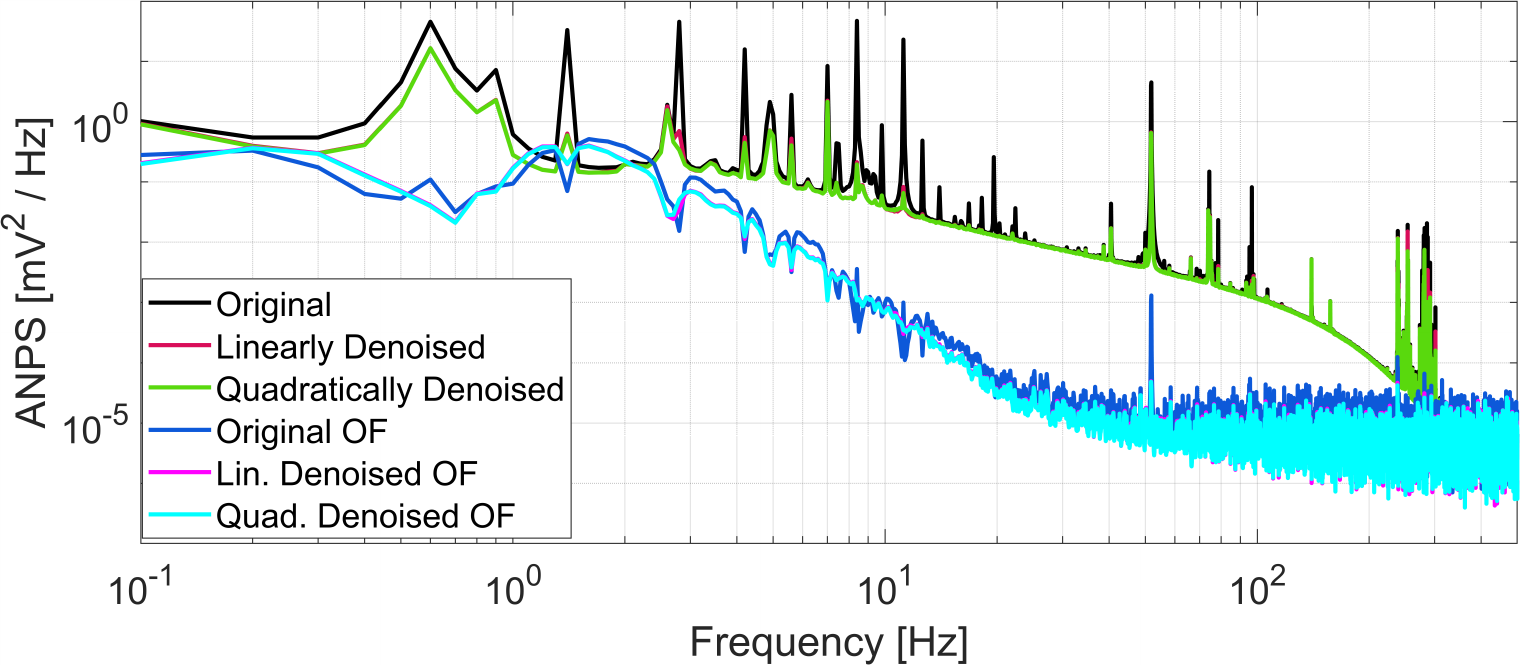}
\caption{Effects of the denoising and the OF on the noise in the CUORE channel. The original and denoised curves (before the OF) are the same as those in Figure \ref{fig:cuore_AP_ANPS}, but with a coarser binning. This is because the denoising algorithm uses 50-second noise windows, while the OF uses 10-second signal and noise windows. Note the reduction of notches at 1.4 Hz and subsequent harmonics when the data are denoised. Since the OF is normalized to ensure unity gain, reducing the notches has the effect of lowering the continuum of the noise, resulting in lower total noise power in the system.} \label{fig:cuore_ANPS}
\end{figure}

To evaluate the performance of the different denoising methods, we first examine the distributions of noise events which are shown in Figure \ref{fig:cuore_allNoiseRMS}. Before the OF is applied, the denoising algorithm reduces the RMS of a typical noise event by more than a factor of 2. We build the OF using a 10-second long signal template. To make the calculation of the OF coefficients as straightforward as possible, we also use 10-second long noise windows to construct the OF as opposed to the 50-second long noise windows we use to perform the denoising. The OF naturally further reduces the RMS noise, but combining the OF with the denoising produces a smaller mean noise event RMS. This value is further reduced when the OF is combined with the nonlinear algorithm. We confirm this effect by building the ANPS of the detector after each signal processing technique is applied (see Figure \ref{fig:cuore_ANPS}). Taking the square root of the integral of each of the ANPS gives the expected RMS detector noise, the values of which are tabulated in Table \ref{Tab:cuore_noise_table}. The denoising improves the noise RMS of the OF signal by 1.0 mV ($\sim$8\%). The quadratic version of the algorithm again out-performs the linear one, improving the noise RMS by 1.2 mV ($\sim$10\%).

\begin{table*}
\begin{center}
\captionof{table}{\label{Tab:cuore_noise_table} Average noise RMS of the CUORE channel after applying different combinations of processing techniques. As expected, the OF alone outperforms the noise decorrelation, but the combination of the OF with the denoising gives the best performance. The noise RMS is minimized when the quadratic denoising algorithm is used along with the OF.}
\begin{tabular}{l | c }
 \hline
 Signal Type & Noise RMS [mV]\\ [0.5ex] 
 \hline
 Raw Signal                       & 7.52 \\
 Linearly Denoised Signal         & 3.16 \\
 Quadratically Denoised Signal    & 3.12 \\
 Optimal Filtered (OF) Signal     & 1.19 \\
 OF Linearly Denoised Signal      & 1.09 \\
 OF Quadratically Denoised Signal & 1.07 \\
 \hline
\end{tabular}
\end{center}
\end{table*}

\begin{figure}[!ht]
\centering
 \includegraphics[width=\linewidth]{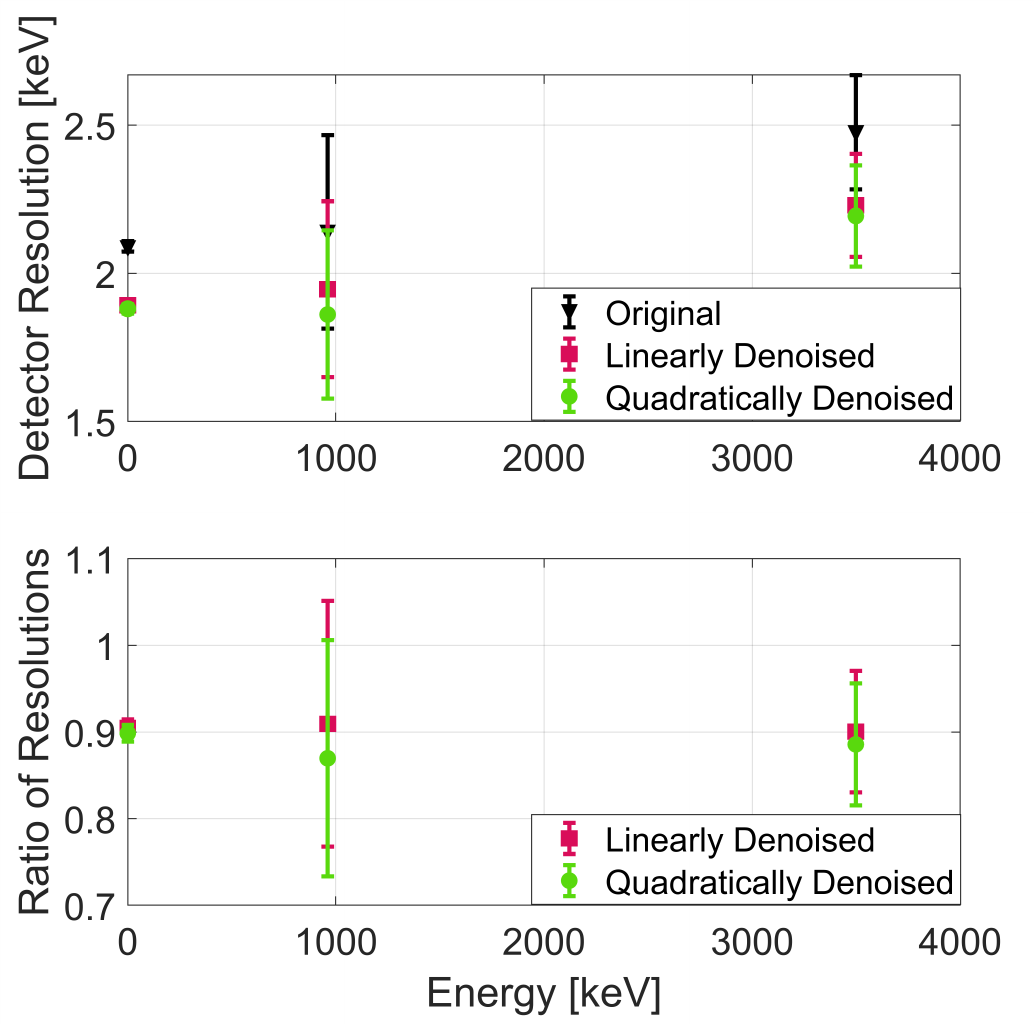}
\caption{Energy resolution of noise events and pulser events before and after the different denoising techniques are applied. The energy resolutions of the pulsers are similar to that of the noise. At all energies, the denoising algorithm improves the resolution of the detector and the nonlinear version further improves the resolution.} \label{fig:cuore_energy_res}
\end{figure}

We now analyze the noise events and pulser events together using the same analysis techniques as described in Section \ref{sec:berk_techniques}. Fitting the OF amplitude spectrum and calibrating using the known pulser energies gives us our final energy spectrum. The obtained resolutions are reported in Table \ref{Tab:cuore_results_table}, while the ratios of resolutions are reported in Table \ref{Tab:cuore_ratio_table}. Looking at the noise peak, where the statistical uncertainty on the resolution is smallest, the relative improvement in resolution is $9\%$ when the linear denoising is applied and $10\%$ when the nonlinear version is applied. This again shows the better performance of the nonlinear method. This $10\%$ improvement in energy resolution is similar to the improvement in the noise RMS, verifying that the calibration method does not qualitatively change the effects of the denoising. The pulser resolutions are comparable to the noise resolution as expected, though the resolution of the 3500 keV pulser is slightly worse than expected. Even so, the relative improvement in energy resolution is compatible at all energies (see Figure \ref{fig:cuore_energy_res}). Together, these results support the hypothesis that the resolution of the pulsers is approximately the same as that of the noise. An energy-dependent contribution to the pulser resolution should not be removed by the denoising algorithm because the auxiliary device signals are completely uncorrelated to the pulse energy. If a non-negligible energy-dependent contribution were present, then the relative change in the resolution of the pulsers after denoising would be different as a function of energy. We observe that the improvements in pulser resolution are compatible at all measured energies, which suggests that the energy-dependent contribution to the pulser energy resolution is negligible. In summary, the denoising algorithm consistently improves the energy resolution of noise events and pulser events by $\sim$10\%. The nonlinear version of the algorithm adds another $\sim$1\% improvement, though this improvement is within the uncertainty of the measurement. 

\begin{table*}
\begin{center}
\captionof{table}{\label{Tab:cuore_results_table} Energy resolution of noise events and pulser events after linear and nonlinear denoising techniques are each applied. The denoising improves the resolution by $\sim$10\%. The quadratic version of the algorithm adds another $\sim$1\% improvement, though this improvement is not statistically significant. Pulser energies were determined during online CUORE data analysis. All energy resolutions are calculated from the optimally filtered and calibrated data.}
\begin{tabular}{l | c c c }
 \hline
 Signal Type & $\hat{\sigma}_{0}$ [keV] & $\hat{\sigma}(E = 960 \text{ keV})$ [keV] & $\hat{\sigma}(E = 3500 \text{ keV})$ [keV]\\ [0.5ex] 
 \hline
 OF Original               & 2.09 $\pm$ 0.02 & 2.1 $\pm$ 0.4 & 2.5 $\pm$ 0.2 \\
 OF Linearly Denoised      & 1.89 $\pm$ 0.02 & 1.9 $\pm$ 0.3 & 2.2 $\pm$ 0.2 \\
 OF Quadratically Denoised & 1.88 $\pm$ 0.02 & 1.9 $\pm$ 0.3 & 2.2 $\pm$ 0.2 \\
 \hline
\end{tabular}
\end{center}
\end{table*}

\begin{table*}
\begin{center}
\captionof{table}{\label{Tab:cuore_ratio_table} Ratio of denoised energy resolution to original energy resolution using linear and nonlinear versions of the algorithm. The denoising algorithm reduces the noise resolution and the pulser energy resolution by compatible amounts. This is expected because the main contribution to the pulser resolution is the noise. The quadratic version of the algorithm further reduces the resolution, though this reduction is not statistically significant. The resulting resolutions are again compatible across all energies.}
\begin{tabular}{l | c c c}
 \hline
 Signal Type & $\hat{R}_{0}$ & $\hat{R}(E = 960 \text{ keV})$ & $\hat{R}(E = 3500 \text{ keV})$ \\ [0.5ex] 
 \hline
 OF Linearly Denoised      & 0.91 $\pm$ 0.01 & 0.91 $\pm$ 0.14 & 0.90 $\pm$ 0.07 \\
 OF Quadratically Denoised & 0.90 $\pm$ 0.01 & 0.87 $\pm$ 0.13 & 0.89 $\pm$ 0.07 \\
 \hline
\end{tabular}
\end{center}
\end{table*}

We can also estimate the timing resolution of the detector before and after the denoising techniques are applied. For a signal of amplitude $A$, the timing resolution of the digital signal is given by \cite{golwala}:

\begin{equation}
    \sigma_{t_0} =  \frac{1}{A} \left[T \sum_{n = -N/2}^{N/2 - 1} (2\pi f_n)^2 \frac{\abs{s(f_n)}^2}{N(f_n)} \right]^{-\frac{1}{2}}
\end{equation}

where $T$ is the length of the signal in seconds, $s(f_n)$ is the Fourier transform of the average pulse, and $N(f_n)$ is the ANPS. The timing resolution is inversely proportional to the amplitude of the signal, and it is a single number for a given average pulse and ANPS. We therefore use this proportionality constant $A\sigma_{t_0}$ as our figure of merit for the different signal processing techniques. We calculate it using the average pulse with amplitude 1 mV and multiply by the value of the calibration function at $A = 1$ mV to report the value in keV$\cdot$s (see Table \ref{Tab:cuore_time_table}). We find that the linear denoising improves the timing resolution by 1.4\% while the quadratic version improves the timing resolution by 1.9\%. We note that the improvement in timing resolution is significantly less than that of the energy resolution. This suggests that in this channel, vibrational noise does not contribute significantly to the timing resolution. 

\begin{table*}
\begin{center}
\captionof{table}{\label{Tab:cuore_time_table} Estimated timing resolution of the CUORE channel after applying different combinations of processing techniques. The estimated timing resolution is minimized when the quadratic denoising algorithm is used along with the OF. The improvement is of order 1\% after the denoising.}
\begin{tabular}{l | c }
 \hline
 Signal Type & Timing Resolution [keV$\cdot$ms]\\ [0.5ex] 
 \hline
 Optimal Filtered (OF) Signal     & 26.91 $\pm$ 0.11 \\
 OF Linearly Denoised Signal      & 26.54 $\pm$ 0.11 \\
 OF Quadratically Denoised Signal & 26.41 $\pm$ 0.11 \\
 \hline
\end{tabular}
\end{center}
\end{table*}

The main limitations of the algorithm are likely related to the auxiliary devices. First, it is important that the auxiliary devices be sensitive to noise in relevant signal bands. The magnitude of the OF applied to the raw timestream peaks near 1.4 Hz, where the PT noise is very well measured. At frequencies below $\sim$1 Hz, the accelerometer responses roll off significantly, thus the accelerometers are not sensitive to the sub-Hz noise observed in the CUORE data. The seismometers used in CUORE are designed to measure sub-Hz noise with better sensitivity, so the algorithm can reduce the sub-Hz noise peaks in the CUORE data. These peaks are reduced by 4 to 5 dB and are far from eliminated. The performance of the denoising is also directly tied to the coherence of the auxiliary devices with the CUORE channel, i.e. the fraction of the power in the auxiliary signals that is correlated with the detector at a particular frequency. It is possible that the coherence of sub-Hz noise between the seismometer signal and detector noise can be increased by moving the seismometers elsewhere on the cryostat, as their locations have not yet been optimized. It is also likely that the algorithm can be improved by introducing auxiliary devices in close proximity to the CUORE detector. This suggests that introducing cryogenic vibration sensors into future cryogenic bolometric experiments is to be investigated further. A basic version of this technique using an NTD vibration sensor was demonstrated in Section \ref{sec:berkeley_setup}.

Finally, it is instructive to compare the effects of this denoising technique with the noise decorrelation technique used in CUORICINO \cite{cuoricino_noise}. In CUORICINO, neighboring bolometer channels were used as auxiliary devices to remove noise that was correlated between them. The CUORICINO technique effectively removed sharp peaks from the noise power spectrum much like the technique presented here. However, the CUORICINO technique was effective at removing continuum noise below 1 Hz, which our technique fails to do. This suggests that the noise measured in the auxiliary devices used in CUORE is not correlated with the continuum noise in the CUORE detector. In \cite{cuoricino_noise}, it was found that most CUORICINO channels do not exhibit more than a 5\% improvement in OF noise resolution, though the largest improvements were more than 50\%. It is unclear whether the CUORE channel showed here is one of the most-improved or least-improved CUORE channels using our noise correlation algorithm, but the 10\% improvement in OF resolution we report here out-performs the vast majority of channels in the CUORICINO analysis.

It should also be noted that the CUORICINO noise decorrelation was done using 11 bolometer channels as auxiliary channels. In practice, this creates a problem due to "side-pulses," i.e. pulses present on neighboring bolometer channels. The presence of side-pulses creates difficulties as one must find a way to remove them, either by fitting the pulses on the auxiliary channels and subtracting them or by removing the channel(s) containing side-pulses and re-computing the transfer functions between the remaining channels. The probability of encountering side-pulses grows as the number of channels used in the decorrelation increases. Furthermore, neighboring bolometer channels are the most likely to have correlated noise, but they are also the most likely to have coincident pulses due to multi-site physics events. The denoising technique presented here does not incur such problems since there are no side-pulses on the auxiliary devices used in CUORE. Furthermore, this technique denoises bolometer channels against auxiliary devices, which has not been attempted before in bolometric experiments, and it introduces nonlinear terms motivated by detector response.

\section{Conclusions}

We presented an algorithm based on a multi-input, single-output model which can be used in a variety of settings to remove noise from detector signals using multiple auxiliary devices such as microphones, accelerometers, and seismometers. The algorithm can be applied to detectors with a wide range of bandwidths and offers an inexpensive way to reduce noise without major experimental interventions. We validated the approach by implementing the multivariate noise decorrelation algorithm on simulated calorimeter data with simulated noise contributions from electromagnetic interference and microphonic vibrations. We then implemented the algorithm on a low-temperature light detector using accelerometers, an antenna, and an NTD vibration sensor as auxiliary devices. The result was a reduction in the noise RMS of the detector and a 4\% improvement in energy resolution at the noise peak. We then presented the case for using a nonlinear algorithm which accounts for quadratic mixing terms from the auxiliary device signals to denoise thermal detectors. We demonstrated the nonlinear version of the algorithm with one channel from the CUORE experiment, showing an improvement in the OF noise RMS and the energy resolution of noise events and pulser events. The denoising algorithm improved these by approximately $10\%$. The nonlinear version of the algorithm outperformed the linear version, adding a further improvement of 1--2\%. The main effect of the denoising was a reduction in noise peaks but there was no clear reduction in the broadband noise. The timing resolution of the channel was also estimated to improve by 1--2\%, with the greatest estimated improvement again coming from the nonlinear version of the algorithm.

Work is ongoing to integrate this denoising algorithm into the analysis framework of CUORE in future searches for rare processes such as neutrinoless double beta decay. Looking ahead, we plan to conduct a rigorous study of the effects of the denoising algorithm, particularly its impact on the timing resolution, energy resolution, and energy threshold of the entire CUORE detector over multiple years of data-taking. This will allow for a better comparison of the denoising technique outlined here with the technique used in CUORICINO. We are also in the process of expanding our array of auxiliary devices in the CUORE experiment by adding antennas and more accelerometers to our DAQ system. Beyond this, we will further investigate the use of bicoherence and higher-order frequency mixing terms to more accurately model the effects of vibrations in thermal detector signals, extending the model proposed in \cite{thermalModel}. This will lay the ground work as we build toward a multivariate nonlinear adaptive filter based on the single-input model demonstrated in \cite{zimmermann2013active}. Active filters such as these have the potential to be integrated into the online data-taking of future experiments including CUPID \cite{cupid}.

\subsection{Acknowledgments} 

The authors would like to thank Dr. Steven M. Kay for his guidance and suggestions regarding the spectral analysis of nonlinear systems as well as Dr. Arnaud All\'ezy for his work on modeling the vibrational modes of the CUORE cryostat. We also thank the CUORE Collaboration for helping to make this work possible. This work was supported by the US Department of Energy (DOE), Office of Science under Contract No. DE-AC02-05CH11231, and by the DOE Office of Science, Office of Nuclear Physics under Contract No. DE-FG02-00ER41138. This work was also supported by the Istituto Nazionale di Fisica Nucleare (INFN). This work makes use of both the DIANA data analysis and APOLLO data acquisition software packages, which were developed by the CUORICINO, CUORE, LUCIFER, and CUPID-0 Collaborations.

\subsection{Data Availability Statement}

The CUORE data used in this paper are a small subset of the much larger CUORE dataset which will be analyzed in future works. The code used to complete this analysis and the raw data analyzed here (including waveform data from the NTD-Ge detector, the CUORE channel, and all auxiliary devices) are not currently publicly available but they can be made available upon request. 

\begin{figure}[!hb]
\begin{subfigure}{\linewidth}
 \includegraphics[width=\linewidth]{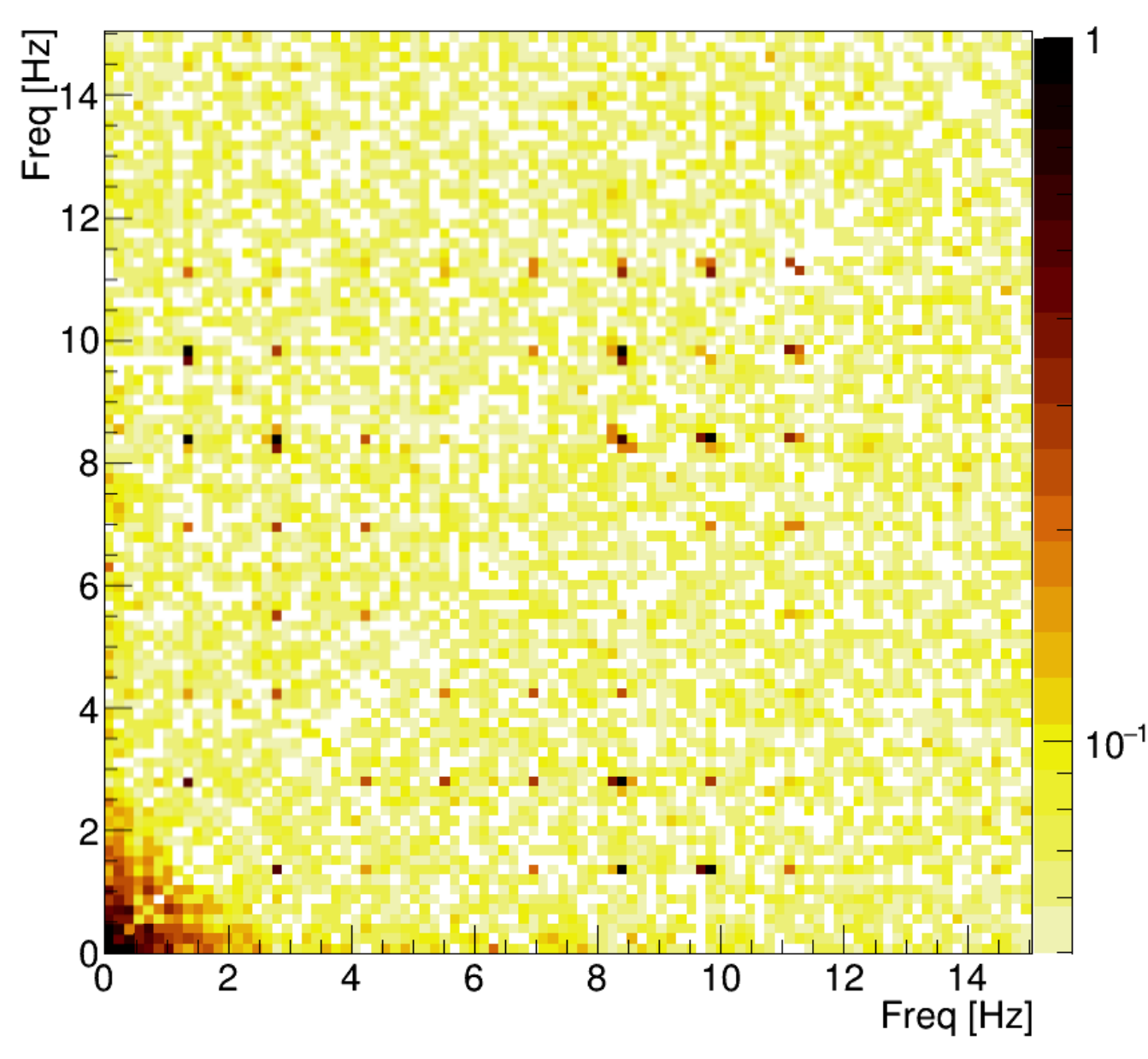}
 \caption{}
\end{subfigure}
\begin{subfigure}{\linewidth}%
 \includegraphics[width=\linewidth]{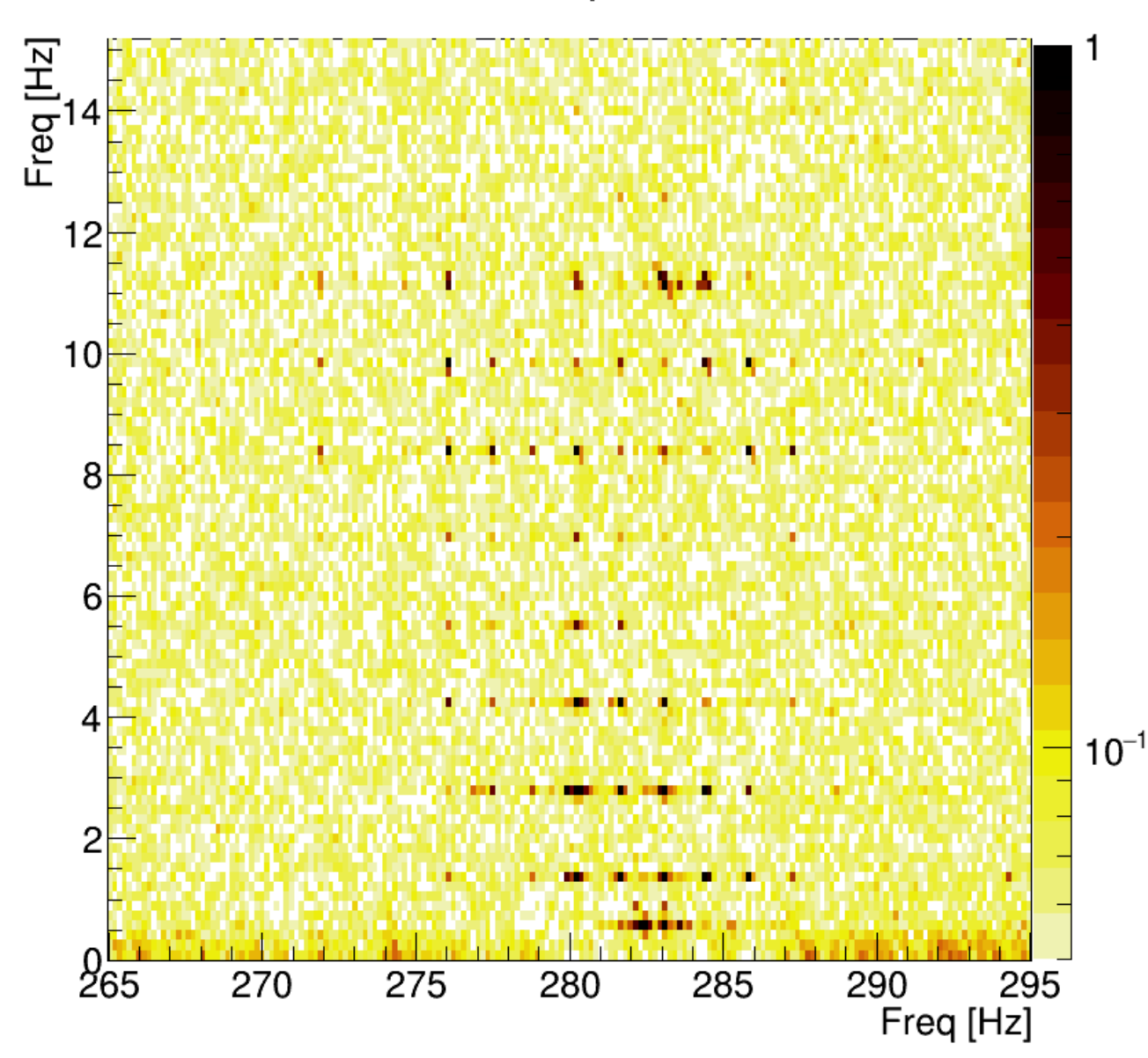}
 \caption{}
\end{subfigure}
\caption{(a) Region of non-zero bicoherence in CUORE data at harmonics of 1.4 Hz induced by the PT phase coupling. (b) Region of non-zero bicoherence in CUORE data indicating mixing of frequencies up to $\sim$12 Hz with frequencies near 280 Hz.}\label{fig:cuore_bicoherence}
\end{figure}

\begin{appendices}

\section{Bicoherence}\label{app:bicoherence}

Using the same notation as in Section \ref{sec:math} where possible, we define the bispectrum of a signal $x[t]$:
\begin{equation}
\mathcal{B}[f_i,f_j] = \expval{X^*[f_i]X^*[f_j]X[f_i + f_j]}
\end{equation}
and we define the bicoherence, $b$ using the following normalization convention:
\begin{equation}
b[f_i,f_j] = \frac{\abs{\mathcal{B}}}{\sqrt{\expval{\abs{X[f_i]X[f_j]}^2}\expval{\abs{X[f_i + f_j]}^2}}}
\end{equation}
where the expectation values are taken over the same noise samples used to compute the denoising transfer function. The bicoherence is restricted to be between 0 and 1, and it measures the proportion of the signal power at any frequency pair that is phase coupled to its sum frequency. This is sometimes interpreted to be a measurement of the nonlinearity present in a system, however there are subtleties related to this. Consider a signal of the form:
\begin{equation}
 x_L(t) = \sum_i A_i \cos(2\pi f_i t + \phi_i(t))
\end{equation}
where $A_i$ are all real coefficients. There is no nonlinear mixing in this system. Now consider three frequencies $f_i, f_j, f_k$ where $f_i + f_j = f_k$. If there is no phase coupling between $f_i$, $f_j$, and $f_k$, then $b^2(f_i, f_j)$ approaches 0 after averaging over a sufficiently long period of time. If the phases are coupled such that over a long time $\expval{\phi_k - (\phi_i  + \phi_j)} \neq 0$, then $b^2(f_i, f_j) > 0$. This phase coupling is expected in systems that are driven by an external periodic source of noise. The PTs in CUORE are an example of such an external source, so we expect non-zero bicoherence at harmonics of the fundamental PT frequency (1.4 Hz) even without the presence of a nonlinear mixing term.

Now we analyze the bicoherence of a system with a quadratic mixing term. Consider a simple signal of containing two frequencies with a nonlinear mixing term. Using $\omega_i = 2\pi f_i$, we have:
\begin{equation}
\begin{split}
x_{NL}(t) &= A_1 \cos(\omega_1t + \phi_1) + A_2 \cos(\omega_2t + \phi_2) \\
& \hspace{10pt} + 2A_{12}\cos(\omega_1 t + \phi_1)\cos(\omega_2 t + \phi_2) \\
 &= A_1 \cos(2\pi \omega_1t + \phi_1) + A_2 \cos(\omega_2t + \phi_2) \\
 & \hspace{10pt} + A_{12}\cos( (\omega_1+\omega_2) t + \phi_1+\phi_2) \\
 & \hspace{10pt} + A_{12}\cos( (\omega_1-\omega_2) t + \phi_1-\phi_2)
\end{split}
\end{equation}

The bicoherence of this signal at $(\omega_1,\omega_2)$ is:
\begin{equation}
 b(f_1,f_2) = \frac{\expval{A_1e^{i\phi_1} A_2e^{i\phi_2} A_{12}e^{-i(\phi_1+\phi_2)}}}{A_1 A_2 A_{12}} = 1
\end{equation}
In this system, all of the power at $f = f_1 + f_2$ is due to the mixing of the signals at $f_1$ and $f_2$, so the bicoherence attains the maximum possible value. 

Analysis of these two signals shows that while a non-zero bicoherence is a natural consequence of a quadratic nonlinearity in a system, it does not guarantee the existence of such a nonlinearity. The bicoherence is therefore useful for examining the phase-coupled components of a signal, but one should be cautious before attributing such phase coupling to an underlying nonlinearity. We discuss why this quadratic mixing model may still be appropriate for the CUORE system in Section \ref{sec:nonlin} and Appendix \ref{app:mix}. 

The bicoherence of a single CUORE channel is plotted in Figure \ref{fig:cuore_bicoherence} for regions of frequency space where the bicoherence is non-negligible. Note that the bicoherence is an auto-quantity; it only depends on the bolometer signal, not a pair of signals. We see a large amount of bicoherence between the different PT frequencies, but we expect the harmonics of the PT phases to be coupled already. An interesting region exists around $f_1 \approx280$ Hz, $f_2 < 10$ Hz. There, we see several points with a large bicoherence term. The distances between these points are integer multiples of 0.6 and 1.4 Hz, both of which appear on their own in the bolometer ANPS. Here, there is no obvious harmonic effect that could contribute to the bicoherence, so a nonlinear mixing term may be at play. We also note that it is at these frequencies near 280 Hz where we see the greatest difference in noise power between the linearly denoised and quadratically denoised signal, as discussed in Section \ref{sec:cuore_results} and shown in Figure \ref{fig:anps_zooms}. This is further evidence that the noise power at these frequencies is due to a nonlinear effect.

\section{Frequency Mixing in Power Signals}\label{app:mix}

Consider an object of mass $m$ vibrating in 1-dimension whose acceleration is described by a discrete signal $a[t]$. The power supplied by this oscillation is equal to $p[t] = ma[t]v[t] = m a[t] \sum_{t' < t} a[t'] dt' $. (We will drop the $m$ so that $p$ has units of $t(a[t])^2$. Taking the Fourier transform of the power, we have:
\begin{equation}\label{eq:power_sum}
 P[\omega] = A[\omega] \circledast \frac{A[\omega]}{\omega} = i \sum_{\omega' \neq \omega}\frac{A[\omega'] A[\Delta \omega']}{\Delta \omega'}
\end{equation}

where $\Delta \omega' = \omega - \omega'$ and $\circledast$ represents the convolution. One can replace $(\Delta \omega')$ with $(2\pi f_{s} + \Delta \omega')$ if $\omega' > \omega$ when working on the non-negative frequency domain. Now consider the square of the acceleration signal $(a[t])^2$. Its Fourier transform is: 
\begin{equation}\label{eq:squaresum}
 A^2[\omega] = \sum_{\omega' \neq \omega}A[\omega'] A[\Delta \omega']
\end{equation}

The transfer function from this signal to the power signal is:
\begin{equation}
 H_{x^2P} = \frac{\expval{\left(A^2[\omega]\right)^*P[\omega]}}{\expval{\left(A^2[\omega]\right)^*\left(A^2[\omega]\right)}}
\end{equation}

Plugging in, we see the transfer function is averaged over all mixing frequencies, the $P[\omega]$ term of which is averaged in a frequency-weighted way:
\begin{equation}\label{eq:TF}
\begin{split}
&H_{x^2P}=\\
&\frac{ i \expval{ \left( \sum_{\omega' } A^*[\omega'] A^*[\Delta \omega'] \right) \left(\sum_{\omega'} \frac{A[\omega'] A[\Delta \omega'] }{\Delta \omega' }\right)}}{\expval{ \abs{\sum_{\omega' } A[\omega'] A[\Delta \omega'] }^2}} \\
=&\frac{ i
\expval{ \sum_{\omega' } \sum_{\omega'' } \frac{ 
A^*[\omega'] A^*[\Delta \omega'] A[\omega''] A[\Delta \omega'']
}{\Delta \omega''}}}{ \expval{ \abs{\sum_{\omega' } A[\omega'] A[\Delta \omega'] }^2}} \\
\end{split}
\end{equation}

The bispectrum $\mathcal{B}^2[\omega', \Delta \omega']$ of the original acceleration signal is now a useful quantity, since it tells us which frequency pairs will contribute to the expectation value of the above sum. If the matrix is sparse, then the number of frequency pairs over which one must sum becomes significantly smaller, and evaluating the sum for each $\omega$ becomes significantly less computationally expensive. Nonetheless, the transfer function $H_{x^2P}$ still provides a ``frequency-smeared'' approximation of the expected value of the power signal given the square of the accelerometer signal. In the limit that a single frequency pair $(\omega_0, \omega - \omega_0)$ dominates the sums in equations \ref{eq:power_sum} and \ref{eq:squaresum}, we can approximate the signals as:
\begin{equation}
\begin{split}
 P[\omega] &= i A[\omega_0] A[\omega - \omega_0]\left( \frac{1}{\omega_0} + \frac{1}{\omega- \omega_0} \right) \\
&= \frac{i\omega A[\omega_0] A[\omega - \omega_0]}{\omega_0 (\omega - \omega_0)} \\
 \end{split} 
 \end{equation}

 \begin{equation}
 A^2[\omega] = 2A[\omega - \omega_0] A[\omega - \omega_0]
\end{equation}

Plugging this into \ref{eq:TF}, the factors of $\expval{(A[\omega_0] A[\omega - \omega_0])^*(A[\omega_0] A[\omega - \omega_0])}$ cancel assuming there is no uncorrelated noise in the acceleration signal at $\omega$. The transfer function is:
\begin{equation}
 H_{x^2P} = \frac{i\omega}{2\omega_0 (\omega - \omega_0)}
\end{equation}

In this case, the transfer function accurately describes the frequency mixing behavior while accounting for the difference in units between $P$ and $(a[t])^2$. Of course, this approximation of a single perfectly bicoherent frequency pair contributing to the sum in \ref{eq:TF} is not valid in the CUORE system since there are multiple frequency pairs exhibiting bicoherence that sum to the same frequency (see Figure \ref{fig:cuore_bicoherence}). Still, using the frequency-smeared coefficients from the squares of the auxiliary signals in the denoising algorithm results in an improvement over the linear version.

\end{appendices}
\clearpage
\bibliographystyle{ieeetr}
\bibliography{citations}

\end{document}